\documentclass[10pt,journal]{IEEEtran} 
\makeatletter
\usepackage{subfigure}
\usepackage{times}
\usepackage{caption}
\usepackage{amsthm}
\newcommand{\subparagraph}{}
\theoremstyle{definition}

\long\def\@makecaption#1#2{\ifx\@captype\@IEEEtablestring%
\footnotesize\begin{center}{\normalfont\footnotesize #1}\\
{\normalfont\footnotesize\scshape #2}\end{center}%
\@IEEEtablecaptionsepspace
\else
\@IEEEfigurecaptionsepspace
\setbox\@tempboxa\hbox{\normalfont\footnotesize {#1.}~~ #2}%
\ifdim \wd\@tempboxa >\hsize%
\setbox\@tempboxa\hbox{\normalfont\footnotesize {#1.}~~ }%
\parbox[t]{\hsize}{\normalfont\footnotesize \noindent\unhbox\@tempboxa#2}%
\else
\hbox to\hsize{\normalfont\footnotesize\hfil\box\@tempboxa\hfil}\fi\fi}
\makeatother
\usepackage{booktabs}
\usepackage{array}
\newcolumntype{P}[1]{>{\centering\arraybackslash}p{#1}}
\usepackage{lscape}
\usepackage{comment}
\usepackage{verbatim}
\usepackage{ltablex}
\usepackage[numbers]{natbib}
\usepackage{lscape}
\usepackage[ruled]{algorithm2e}
\usepackage{wrapfig}
\usepackage{color,soul}
\usepackage{longtable}
\usepackage{amssymb,amsmath}
\usepackage{multirow}
\usepackage{varwidth}
\usepackage{colortbl}
\usepackage[usenames,dvipsnames]{xcolor}
\usepackage[pdftex]{graphicx}
\usepackage{stfloats}
\usepackage{graphicx}
\usepackage{tikz}
\usetikzlibrary{er}
\usetikzlibrary{shapes}
\usetikzlibrary{shapes.gates.logic.US,trees,positioning,arrows}
\usepackage{pgfplots}
\pgfplotsset{compat=1.17}
\pgfplotsset{width=18cm, height=6cm}
\usepackage{pifont}

\usepackage[normalem]{ulem}
\usepackage{array,etoolbox}
\preto\tabular{\setcounter{magicrownumbers}{0}}
\newcounter{magicrownumbers}

\usepackage[switch,columnwise]{lineno}
\usepackage{pgf-pie}
\SetAlFnt{\small}
\SetAlCapFnt{\small}
\SetAlCapNameFnt{\small}
\SetAlCapHSkip{0pt}
\IncMargin{-\parindent}

\tikzstyle{chart}=
[legend label/.style={font={\scriptsize},anchor=west,align=left},
legend box/.style={rectangle, draw, minimum size=5pt},
axis/.style={black,thin,->},
axis label/.style={anchor=east,font={\tiny}}]

\tikzstyle{bar chart}=[
chart, bar width/.code={
    \pgfmathparse{##1/2}
    \global\let\bar@w\pgfmathresult},
bar/.style={thin, draw=black},
bar label/.style={font={\bf\small},anchor=north},
bar value/.style={font={\footnotesize}},
bar width=.75,]

\tikzstyle{pie chart}=
[chart,
slice/.style={line cap=round, line join=round, thin, draw=black},
pie title/.style={font={}},
slice type/.style 2 args={
    ##1/.style={fill=##2},
    values of ##1/.style={}}]

\pgfdeclarelayer{background}
\pgfdeclarelayer{foreground}
\pgfsetlayers{background,main,foreground}

\newcommand{\tempsum}{0}

\newcommand{\pye}[3][]{
    \begin{scope}[#1]
        \pgfmathsetmacro{\curA}{90}
        \pgfmathsetmacro{\r}{1}
        \def\c{(0,0)}
        \def\tempsum{0}
        \foreach \v/\s in{#3}{
          \pgfmathparse{\v+\tempsum}
          \global\let\tempsum=\pgfmathresult}
        \foreach \v/\s in{#3}{
            \pgfmathsetmacro{\deltaA}{\v/\tempsum*360}
            \pgfmathsetmacro{\nextA}{\curA + \deltaA}
            \pgfmathsetmacro{\midA}{(\curA+\nextA)/2}
            \path[slice,\s] \c
            -- +(\curA:\r)
            arc (\curA:\nextA:\r)
            -- cycle;
            \pgfmathsetmacro{\d}{1.2}
            \begin{pgfonlayer}{foreground}
                \path \c -- node[pos=\d,pie values,values of \s]{$\v$} +  (\midA:\r);
            \end{pgfonlayer}
            \global\let\curA\nextA}
    \end{scope}}

\newcommand{\legend}[2][]{
    \begin{scope}[#1]
        \path
        \foreach \n/\s in {#2}
        {++(0,-5pt) node[\s,legend box] {} +(2pt,0) node[legend label]     {\n}};
    \end{scope}}
\usepackage{pgfplots}
\definecolor{ac1}{HTML}{b88b4d}

\newcommand{\new}{\textcolor{black}}

\newcommand{\jhc}[1]{\textcolor{red} {#1}}

\usepackage{enumitem}
\setlist[itemize]{leftmargin=*}

\newcommand{\fully}{\ding{52}}
\newcommand{\partly}{\ding{115}}
\newcommand{\notso}{\ding{56}}

\begin{document}

\title{Game-Theoretic and Machine Learning-based Approaches for Defensive Deception: A Survey}

\author{Mu Zhu, Ahmed H. Anwar, Zelin Wan, Jin-Hee Cho, Charles Kamhoua, and Munindar P. Singh
\IEEEcompsocitemizethanks{\IEEEcompsocthanksitem Zelin Wan and Jin-Hee Cho are with the Department of Computer Science, Virginia Tech, Falls Church, VA, USA. Email: \{zelin, jicho\}@vt.edu. Email: abdullahzubair@vt.edu. Mu Zhu and Munindar P. Singh are with the Department of Computer Science, North Carolina State University, Raleigh, NC 27695. Email: \{mzhu5, mpsingh\}@ncsu.edu. Ahmed H. Anwar and Charles A. Kamhoua are with the US Army Research Laboratory, Adelphi, MD, USA. Email: a.h.anwar@knights.ucf.edu; charles.a.kamhoua.civ@mail.mil.}}

\newpage

\maketitle

\begin{abstract}
\new{Defensive deception is a promising approach for cyber defense. Via defensive deception, the defender can anticipate attacker actions; it can mislead or lure attacker, or hide real resources. Although defensive deception is increasingly popular in the research community, there has not been a systematic investigation of its key components, the underlying principles, and its tradeoffs in various problem settings. This survey paper focuses on defensive deception research centered on game theory and machine learning, since these are prominent families of artificial intelligence approaches that are widely employed in defensive deception. This paper brings forth insights, lessons, and limitations from prior work. It closes with an outline of some research directions to tackle major gaps in current defensive deception research.}
\end{abstract}

\IEEEpeerreviewmaketitle

\section{Introduction}
\label{sec:intro}


\subsection{Motivation}
\label{subsec:motivation}

Conventional security mechanisms, such as access controls and intrusion detection, help deal with outside and inside threats but inadequately resist attackers subverting controls or posing new attacks. Deception is a distinct line of defense aiming to thwart potential attackers. The key idea of deception is to manipulate an attacker's beliefs to mislead their decision making, inducing them to act suboptimally. Since the benefits of leveraging the core ideas of defensive deception have been realized in the cybersecurity research community, there have been non-trivial efforts to develop intelligent defensive deception techniques. 

Two main promising directions to develop defensive deception techniques are observed in the literature. First, strategies of an attacker and defender have been commonly modeled based on game-theoretic approaches where the defender takes defensive deception strategies with the aim of creating confusion for attackers or misleading them to choose less optimal or poor strategies. Second, machine learning (ML)-based defensive deception techniques have been proposed to create decoy objects or fake information that mimic real objects or information to mislead or lure attackers.

\new{The synergistic merit of combining GT and ML has been recognized in the cybersecurity literature~\cite{kamhoua2021game}, such as using game theoretic defenses against adversarial machine learning attacks~\cite{Dasgupta19, Zhou19} or generative adversarial models for creating deceptive objects~\cite{kamhoua2021game}.  However, little work has explored the synergies between GT and ML to formulate various cybersecurity problems. In particular, since players' effective learning of their opponents' behavior is critical to the accuracy of their beliefs of the opponents' types or next moves, using ML for the players to form their beliefs can contribute to generating optimal plays under a certain environment. In addition, when developing defensive deception techniques, ML-based approaches can provide better prediction of attackers or high similarity in creating deceptive object based on a large volume of data available. However, they may not provide effective strategic solutions under uncertainty which has been well explored in game theoretic approaches.  Therefore, this survey paper was motivated to facilitate future research taking hybrid defensive deception approaches that can leverage both GT and ML.}

In order to distinguish the key contributions of our paper compared to the existing survey papers, we discuss the existing survey papers on defensive deception techniques and clarify the differences between our paper and them in the following section. 

\begin{table*}[t]
\centering
\scriptsize
\caption{Comparison of our Survey Paper with the Existing Surveys of Defensive Deception}
\label{tab:Comparison}
\vspace{-7mm}
\begin{center}
\begin{tabular}{|p{5cm}|P{1.4cm}|P{1.7cm}|P{1.7cm}|P{2.1cm}|P{1.7cm}|P{1.9cm}|}
\hline
\multicolumn{1}{|c|}{\bf Key Criteria} & {\bf Our Survey (2020)} &{\bf \citet{lu2020cyber} (2020)}& {\bf \citet{pawlick2019game} (2019)} &  {\bf \citet{han2018CSUR} (2018)} & {\bf \citet{rowe2016introduction} (2016)}& {\bf \citet{almeshekah2014planning} (2014)}\\
\hline
Provides concepts & \fully &\fully& \fully  & \fully & \fully & \fully\\
\hline
Provides key taxonomies & \fully &\fully& \fully&\fully & \fully&\fully \\
\hline
Includes ML approaches & \fully & \notso& \notso & \notso & \notso & \notso \\
\hline
Includes game-theoretic approaches & \fully & \partly& \fully & \partly & \partly  & \notso\\
\hline
Describes attack types countermeasured & \fully & \notso& \partly & \partly & \fully & \partly\\
\hline
Describes metrics  & \fully &\notso & \partly& \partly &\partly  & \partly \\
\hline
Describes evaluation testbeds & \fully & \notso&\notso  & \partly &\partly& \partly \\
\hline
Is not limited to specific application domains 
& \fully &\partly& \notso & \fully 
& \fully& \partly \\
\hline
Discusses pros and cons of relevant techniques & \fully &\partly & \partly  & \fully & \fully & \partly \\
\hline
Discusses insights, lessons, and limitations & \fully & \partly&\partly  & \fully & \fully & \partly\\
\hline
Discusses future research directions & \fully &\fully& \fully& \fully & \partly & \notso  \\
\hline
\end{tabular}
\fully: Fully addressed; \partly: Partially addressed; \notso: Not addressed at all.
\end{center}
\vspace{-5mm}
\end{table*}

\subsection{Comparison with Existing Surveys}
\label{subsec:similar-survey-papers}

Several studies have conducted surveys of defensive deception techniques \cite{lu2020cyber, pawlick2019game,han2018CSUR,rowe2016introduction,almeshekah2014planning}. 

\citet{almeshekah2014planning} presented how a defensive deception has been considered in the cyber security defense domain. To be specific, the authors discussed the following three phrases in considering a defensive deception technique: planning, implementing and integrating, and monitoring and evaluating. In particular, this paper discussed the models of planning deceptions in terms of affecting an attacker's perception which can be misled for a defender to achieve a system's security goals. However, this survey paper is limited its contribution to modeling and integrating defensive deception to a limited set of attackers. In addition, this work \new{did not} consider a variety of network environments which should be considered in implementing defensive deception techniques.

\citet{rowe2016introduction} classified defensive deception techniques in terms of impersonation, delays, fakes, camouflage, false excuses, and social engineering. They not only introduced the background on deception technologies but also explored the calculation of detectability and effectiveness of defensive deception. However, their survey of game-theoretic defensive deception is limited and lacks discussion of the state-of-the-art techniques.

\citet{han2018CSUR} surveyed defensive deception techniques based on four criteria, including the goal, unit, layer, and deployment of deception. They surveyed theoretical models used for defensive deception techniques as well as the generation, placement, deployment, and monitoring of deception elements. \citeauthor{han2018CSUR} discussed the tradeoffs between various deceptive techniques in relation to whether they are deployed at the network, system, application, or data layers. Their discussion of game-theoretic deception, however, is not comprehensive.

\citet{pawlick2019game} conducted an extensive survey on defensive deception taxonomies and game-theoretic defensive deception techniques that have been used for cybersecurity and privacy. The authors discussed the main six different types of deception categories: perturbation, moving target defense, obfuscation, mixing, honey-X, and attacker engagement. Their paper surveyed 24 papers published over 2008--2018, and defined related taxonomies to develop their own classification of game-theoretic defensive deception techniques. This work is interesting to treat moving target defense and obfuscation as subcategories under defensive deception. This paper discussed the common game-theoretic approaches used for developing defensive deception techniques, such as Stackelberg, Nash, and signaling game theories. However, the survey and analysis of the existing game-theoretic defensive techniques conducted in this paper are limited to game-theoretical analysis without considering realistic network environments where ML-based defensive deception techniques or combination of these two (i.e., game theory and ML) may provide more useful insights and promising research directions. 

Recently, \citet{lu2020cyber} conducted a brief survey based on the processes of defensive deception consisting of three phases: planning of deception, implementation and deployment of deception, and monitoring and evaluation of deception. The authors discussed deception techniques based on information dissimulation to hide real information and information simulation to focus on attackers. This work briefly discussed game-theoretic defensive deception and mainly focused on discussing challenges and limitations of the current research. However, only a small fraction of the literature was included. In addition, this paper \new{did not} discuss ML-based defensive deception approaches.

Some survey papers mainly focused on defensive deception techniques for a particular type of attacks or particular deception techniques. 
\citet{carroll2011game} investigated the effects of deception on the game-theoretic interactions between an attacker and defender of a computer network. They examined signaling games and related Nash equilibrium. However, this investigation only focused on honeypots technology while game-theoretic analysis of the deception is limited in examining the interactions between an attacker and a defender in the signaling games.  \citet{virvilis2014changing} surveyed partial defensive deception techniques that can be used to mitigate Advanced Persistent Threats (APTs). 

In Table~\ref{tab:Comparison}, we summarized the key contributions of our survey paper, compared to those of the five existing survey papers~\cite{lu2020cyber, pawlick2019game,han2018CSUR,rowe2016introduction,almeshekah2014planning} based on several key criteria.

\subsection{Key Contributions} \label{subsec:key-contributions}
In this paper, we made the following {\bf key contributions}:
\begin{enumerate}[leftmargin=*]
\item We provided a novel classification scheme that characterizes a defensive deception technique in terms of its conceptual deception categories, presence of an object (i.e., physical or virtual), expected effects after applying deception, ultimate goal (i.e., for asset protection or attack detection), and activeness (i.e., active or passive or both). This provides an in-depth understanding of each deception technique and insights on how it can be applied to support a system's security goal.
\item We discussed key design principles of defensive deception techniques in terms of what-attacker-to-deceive, when-to-deceive, and how-to-deceive. In addition, based on the key properties of defensive deception techniques, we identified the key benefits and caveats when developing defensive deception techniques leveraging game theory and ML algorithms.
\item We discussed game-theoretic and ML-based defensive deception techniques based their types along with pros and cons. In addition, using the classification scheme we introduced in Section~\ref{sec:taxonomies}, we discussed both game-theoretic and ML algorithms.
\item We also surveyed attacks that are handled by the existing game-theoretic and ML-based defensive deception techniques. Accordingly, we discussed what attacks are more or less considered in the literature by the defensive deception techniques.
\item We surveyed how defensive deception techniques are mainly considered to deal with the challenges of different network environments as application domains and discussed pros and cons of the deployed game-theoretic or ML-based defensive deception techniques.
\item We examined what types of metrics and experiment testbeds are more or less used in game-theoretic or ML-based defensive deception techniques to prove their effectiveness and efficiency.
\item We extensively discussed lessons and insights learned and limitations observed from the defensive deception techniques surveyed in this work. Based on these insights and limitations learned, we suggested promising future directions for game-theoretic and ML-based defensive deception research. 
\end{enumerate}
Note that the scope of this paper is mainly focused on surveying game-theoretic (GT) or ML-based defensive deception techniques and discussing insights, limitations, or lessons learned from this extensive survey. Hence, some defensive deception techniques that are not using either game-theoretic approaches or ML are excluded in this survey paper.

\subsection{Research Questions} \label{subsec:research-questions}
We address the following {\bf research questions} in this paper.
\begin{description}[leftmargin=3mm]
\item [RQ Characteristics:] What key characteristics of defensive deception distinguish it from other defensive techniques?
\item [RQ Metrics:] What metrics are more or less used to measure the effectiveness and efficiency of the existing game-theoretic or ML-based defensive deception techniques? 
\item [RQ Principles:] What key design principles help maximize the effectiveness and efficiency of defensive deception techniques?
\item [RQ GT:] What are the key design features when a defensive deception technique is devised using game theory (GT)?
\item [RQ ML:] What are the key design features when a defensive deception technique is developed using ML?
\item [RQ Applications:] How should different defensive deception techniques be applied in different application domains?
\end{description}

We answered these questions in Section~\ref{subsec:insights-lessons}.

\subsection{Structure of the Paper} \label{subsec:paper-structure}

The rest of this paper is structured as follows:
\begin{itemize}[leftmargin=*]
\item Section~\ref{sec:taxonomies} provides the concept of deception and taxonomies related to defensive deception.
\item Section~\ref{sec:design-principles} discusses the key principles of designing a defensive deception technique. In addition, this section clarifies the key distinctive characteristics of the defensive deception techniques, compared to other defense techniques that achieve a same defense goal.
\item Section~\ref{sec:game-dd} explains the key components in using game-theoretic defensive deception and surveys the existing game-theoretic defensive deception techniques along with the discussions of their pros and cons. 
\item Section~\ref{sec:ml-dd} discusses the key components in leveraging ML techniques to develop defensive deception techniques. In addition, this section extensively surveys the existing ML-based defensive deception techniques and addresses their pros and cons. 
\item Section~\ref{sec:attacks} describes attack types countered by the existing game-theoretic and ML-based defensive deception techniques.
\item Section~\ref{sec:evaluation} presents metrics to measure effectiveness and efficiency of the existing defensive deception techniques using game theory and ML. In addition, this section surveys evaluation testbeds used for validating those existing defensive deception techniques surveyed in this work.
\item Section~\ref{sec:app-domains} discusses how game theoretic or ML-based defensive deception techniques have been developed for different application domains, such as enterprise networks, cyberphysical systems (CPS), cloud web-based networks, Internet of Things (IoT), software-defined networks (SDNs), and wireless networks.
\item Section~\ref{sec:conclusions} summarizes the insights and lessons learned by answering the key research questions raised in Section~\ref{subsec:research-questions}. In addition, this section discusses the limitations found from the defensive deception techniques surveyed in this work and suggests promising future research directions.
\end{itemize}

\section{Taxonomies of Defensive Deception} \label{sec:taxonomies}
Deception has been heavily used by attackers which perform a variety of attacks at different levels of applications in both computer and social systems. In this section, we limit our discussions of deception in terms of a defender's perspective. \new{This section mainly includes the discussions of formal models of deception, related common taxonomies used in developing defensive deception techniques, and their distinctive characteristics.} 

\subsection{\new{Formal Models of Cyber Deception}} \label{subsec:formal-deception-model}

\new{
In this section, we discuss formal models of cyber deception that have been discussed in the literature. Although there are numerous formal models of cyber deception, we limit our discussion to formal modeling approaches based on logic, probabilistic logic, and hypergame theory.}

\subsubsection{\new{Logic-based Cyber Deception}}
\new{Deception planning has been studied based on logical reasoning of the key components of deception and their states (i.e., status). 
\citet{Jafarian20-formal-model-deception} and \citet{takabi17-formal-model-deception} proposed deception modeling logic that provides a formal deception model, $\mathcal{DM}$, as:
\begin{gather}
\mathcal{DM} =  
\{\textit{facts}, \textit{beliefs}, \textit{actions}, \\ \textit{causation rules}, \textit{attackers}, \textit{goal}, \textit{budget}\} \nonumber 
\end{gather}
Each component of the DM is:
\begin{itemize}
\item {\em Facts}: The facts indicate a deception problem that can be described by (1) {\em attributes}  of a system (e.g., configuration parameters, such as an operation system of a host, a server's criticality, existing routes between hosts), (2) data types of each attribute (e.g., Boolean, ordered, integral), and (3) actual values assigned to the attributes.  
\item {\em Beliefs}: A belief indicates an attacker's beliefs on the set of attributes, which may not be aligned with the true states of the attributes.
\item {\em Actions}: An action refers to a defender's action to manipulate the attacker's belief on an attribute. Attributes can be {\em actionable} or {\em derivative}. Actionable attributes can be directly manipulated to influence the attacker's belief by taking a deceptive action. Derivable attributes cannot be directly manipulated but can be indirectly derived from the attacker's beliefs over other attributes. When modeling a set of actions, each action should be plausible to the attacker so the deception action is promising enough to deceive the attacker. In addition, the action should consider how much it costs.  
\item {\em Causation rules}: The causation rules apply to each attribute to determine what (e.g., facts or actions) can cause the manipulation of an attacker's belief on the attribute. 
\item {\em Attackers}: A set of attacker types can be modeled based on the probability distribution over the set when the probability information of an attacker being a particular type is available.
\item {\em Goal \& Budget}: The concrete benefit of successful or failed deception should be estimated in terms of defense effectiveness, defense cost (including both monetary budget), expected effect (i.e., a required level of achieved system performance/security) or other associated impact (e.g., interoperability with other defense or service mechanisms).
\end{itemize}
\citet{rowe2007finding} proposed a deception planning method to systematically provide logically consistent deceptions when users are detected as suspicious or malicious by an intrusion detection system. \citet{rowe2007finding} proposed an attack modeling technique that can consider realistic attackers characterized by their targets (e.g., files, a network, status such as log-in or administrator privileges), the status of the targets (e.g.,  existence, authorization, readiness, or operability), and their parameters (e.g., file size, network bandwidth, a number of sites, or a length of the password). By using an inference rule based on the status, an attacker can derive the status of a targeted resource. To prevent the attackers from not achieving their goal, deception tactics can be used to deny the service the attacker requested by generating error messages, such as `a file not transferable', `a file not recognizable', `protection violation', or requesting providing credentials (e.g., password or other credential information).
}

\subsubsection{\new{Probabilistic Cyber Deception}} \new{This type of formal models mainly studies how a developed defensive deception technique can manipulate an attacker's belief and accordingly influence its action based on its belief in a probabilistic manner.  \citet{Jajodia17-prob-logic-deception} proposed a probabilistic logic for cyber deception where a defender sends true or fake results for the scans by an attacker to maximize the probability of successful deception (i.e., maximum manipulation of the attacker's beliefs) while measuring its effectiveness in terms of the probability distribution of the attacker's belief on system attributes (i.e., how much information the attacker has obtained towards a target system accurately). \citet{crouse2015probabilistic} proposed probabilistic models of reconnaissance-based defense with the aim of providing in-depth understanding on the effect of the defense on system security. The authors quantified attack success under various network conditions in terms of a network size, a size of deployment, and a number of vulnerable system components. The proposed deception models provide performance analysis of probabilistic triggering of network address shuffling and deployment of honeypots when these defense mechanisms are considered separately or in concert. In prior work~\cite{Cho19-hgt}, we used Stochastic Petri Nets to model an attack-defense hypergame with the aim of examining how an attacker's beliefs manipulated by a defensive deception can affect its decision making, resulting in attack failure. As \citet{Jajodia17-prob-logic-deception,crouse2015probabilistic} and \citet{ Cho19-hgt} observe, probabilistic models of deception games can effectively capture dynamic interactions between an attacker and a defender under uncertain environments. } 

\subsubsection{\new{Cyber Deception Hypergames}} \new{Much defensive deception research uses game theory to formulate an attack-defense game. In particular, to effectively deal with uncertainty in real scenarios, a sequential attack-defense game where the defender uses defensive deception has been formulated based on hypergame theory. \citet{ferguson2019game} formulated a cyber deception game as a sequential game as $G=(\mathcal{P}, \mathcal{M}, \Theta, u, \mathcal{T})$, where $\mathcal{P}$ refers to a set of players, $\mathcal{M}$ refers to a set of actions, such as $\mathcal{M}=\{\mathcal{M}_i\}$ for player $i$, $\Theta$ is a set of strategies, $\Theta = \{\Theta_i\}$ for player $i$, $u$ is a utility function, $u=\{u_i\}$ for player $i$, and $\mathcal{T}$ is a sequence of moves when players take turns in sequence. By using the formulation of a regular, sequential game, a cyber deception can be formulated as $(G, G^A, G^D)$ where $G^A$ and $G^D$ are derived games. In $G$, a move history of an attacker, $\mathcal{M}^A=\{m^A\}=\{m^A_1, m^A_2, \ldots, m^A_r\}$, and a defender, $\mathcal{M}^D=\{m^D\}=\{m^D_1, m^D_2, \ldots, m^D_s\}$, respectively, refers to the sequence of moves in the game where $r+s \leq \mathcal{T}$. For simplicity, the move histories of the attacker and defender can be denoted by $m=\{m^A, m^D\}$.  Given that $\mathcal{M}^{X|Y}=m^{X|Y}$ refers to $Y$'s perception of the set of $X$'s actions (i.e., it is not necessarily true for $\mathcal{M}^{X|Y} = \mathcal{M}^X$), the attacker (A)'s beliefs about the move sequence, $m$, can be represented by $m^{*|A}=\{m^{A|A}, m^{D|A}\}$. Similar to the notation of the moves perceived by A, we can also obtain $u^{*|A}=(u^A|A, u^D|A)$, which is A's perception of $u=\{u^A, u^D\}$ and $\Theta^{*|A}=\{\Theta^{A|A}, \Theta^{D|A}\}$, which is A's perception of sets of strategies for $\Theta^A$ and $\Theta^D$. Hence, the derived game can be denoted by $G^A=\{\mathcal{P}, \mathcal{M}^{*|A}, u^{*|A}, \mathcal{T}\}$, which is A's perception towards the game $G$, and $G^D=\{\mathcal{P}, \mathcal{M}^{*|D}, u^{*|D}, \mathcal{T}\}$, which is D's perception towards the game $G$. If an attack-defense game does not use the concept of hypergame, it simply relies on the game formulation of $G$.}

\subsection{Concepts of Defensive Deception} \label{subsec:deception-concepts}

The conventional concept of military deception refers to actions taken to intentionally mislead an enemy about one's strengths and weaknesses, intents, and tactics~\cite{sharp2006MilitaryDeception}. The defender employs deception to manipulate the enemy's actions to advance one's mission~\cite{sharp2006MilitaryDeception}. Defensive deception  applies to a wide spectrum of interactions between an attacker and a defender under conflict situations~\citep{sharp2006MilitaryDeception}. The concept of defensive deception initiated in military environments has been introduced to cyber domains with the name of cyberdeception. \citet{almeshekah2016cyber} refined the concept of cyberdeception by \citet{Yuill06} as ``planned actions taken to mislead and/or confuse attackers and to thereby cause them to take (or not take) specific actions that aid computer-security defenses.''  Although the concept of deception is highly multidisciplinary~\cite{Guo20-survey}, the common idea behind deception is agreed as a way to mislead an entity to form a false belief and control its behavior based on it~\cite{rowe2016introduction}. Therefore, when deception is successfully executed, a deceivee responds suboptimally, which provides benefits for a deceiver and results in achieving the deceiver's goal. \citet{rowe2016introduction} emphasized `manipulation' as the core concept of deception where deception encourages a deceivee to do something a deceiver wants or discourages the deceivee from doing something the deceiver does not want.

As the discussion of deception in this work is limited to defensive deception, we consider a defender's deception against an attacker to achieve the defender's goal.

\begin{table*}[th!]
\centering
\caption{Taxonomies and Classification of Defensive Deception Techniques}
\label{tab:technique-effect-goal}
\vspace{-2mm}
\begin{tabular}{|P{2.5cm}|P{3.5cm}|P{4cm}|P{3cm}|P{3cm}|}
\hline
{\bf Deception Tactic} & {\bf Presence of a Real Object} & {\bf Expected Effect} & {\bf Ultimate Goal} & {\bf Activeness} \\
\hline
\hline
Masking & True & Blending, Hiding, Misleading & Asset Protection & Passive \\
\hline
Repackaging & True & Blending, Hiding, Misleading & Asset Protection & Passive \\
\hline
 Dazzling & True & Hiding, Confusing, Misleading & Asset Protection & Passive \\
\hline 
Mimicking & False & Luring, Misleading & Attack Detection & Passive \\
\hline 
Inventing & False & Luring, Misleading & Attack Detection & Active \\
\hline 
Decoying & True & Luring, Misleading & Attack Detection & Active \\
\hline 
Bait & True &  Luring, Misleading & Attack Detection & Active \\
\hline 
Camouflaging & True & Blending, Hiding, Misleading & Asset Protection & Passive \\
\hline 
Concealment & True & Hiding, Misleading & Attack Detection & Active \\
\hline
False Information & False & Luring, Confusing, Misleading & Asset Protection or Attack Detection & Active or Passive \\
\hline
Lies & False & Hiding, Misleading & Asset Protection & Passive \\
\hline
Display & True & Hiding, Misleading & Asset Protection & Passive \\
\hline
\end{tabular}
\vspace{-5mm}
\end{table*}
\subsection{Taxonomies of Defensive Deception} \label{subsec:taxonomies-defensive-deception}

\new{Deception can be applied with certain intent, either malicious or non-malicious intent~\cite{guo2020online}. In particular, deception with malicious intent has been used as an attacker's strategy to deceive a defense system. For example, spammers are paid from clicking deceptive advertisements~\cite{Nextgate13}, malicious users disseminate phishing links to obtain credentials from victim users~\cite{Vergelis19}, or social and political bots propagate false information to influence public opinions~\cite{forelle2015political}.}

As we consider the concept of deception in the context of defensive deception, we limit our discussions of deception `with intent' where the intent is to defend a given system against attackers. In this section, we discuss taxonomies in terms of the following aspects: (i) what conceptual techniques are used; (ii) whether a given deception uses a true object or not (i.e., \new{a} true object exists but is hidden to deceive or a fake object is created to deceive); (iii) what effects are expected when successfully deceiving an opponent; (iv) what is an ultimate goal of the deception, such as detecting an attacker or protecting a system; and (v) whether a given deception is active (mainly for attack detection) or passive (mainly for attack protection) to deceive an opponent.

\subsubsection{Conceptual Deception Tactics}  In the literature~\cite{almeshekah2016cyber, bell1991cheating, bennett2007counterdeception, dunnigan2001victory}, we found the following conceptual deception techniques:
\begin{itemize}[leftmargin=*]
\item {\em Masking}~\cite{almeshekah2016cyber, bell1991cheating}: This refers to hiding certain information or data in the background. 

\item {\em Repackaging}~\cite{almeshekah2016cyber, bell1991cheating}: This means to hide a real object as something else to hide the real object. A defender can make a vulnerability look like something else. For example, a defender can create honey files~\cite{yuill2004honeyfiles}, such as creating trap files that look like regular files.

\item {\em Dazzling}~\cite{almeshekah2016cyber, bell1991cheating}: This is a different way to hide something that is hard to blend in with the background or to be repackaged as something else. A real object can be hidden by overshadowing it, aiming to be undetected as the real, true object. For example, an intruder can be confused by receiving many error messages.

\item {\em Mimicking}~\cite{almeshekah2016cyber, bell1991cheating}: This refers to imitating aspects of a real object. This is often used to look honeypots as real nodes or interfaces to attackers. 

\item {\em Inventing}~\cite{almeshekah2016cyber, bell1991cheating}: A defender can create a new fake object to lure attackers. For instance, a software can be presented in a honeypot for attackers to download it, which allows collecting the attackers' personal data.

\item {\em Decoying}~\cite{almeshekah2016cyber, bell1991cheating}: A deceiver attracts an attacker's attention away from critical assets that should be protected in a system. For example, the deceiver can provide publicly available false information about the system configurations to hinder reconnaissance attacks.

\item {\em Bait}: Deception can use truth~\cite{bennett2007counterdeception} to earn trust from a deceiver. That is, a defender can present attackers with correct but low-grade information to lure them. The truth can be the information that an attacker already has, or the deceiver needs to sacrifice some correct sensitive data to effectively lure attackers. 
\end{itemize}

The military literature provides the following taxonomy for cyberdeception \cite{dunnigan2001victory}. 

\begin{itemize}
\item {\em Concealment}~\cite{dunnigan2001victory}: This deception is  similar to hiding in the classical taxonomy. For the defensive purpose, it helps for honeypots to conceal their user-monitoring software so they look more like normal machines.
    
\item {\em Camouflage}~\cite{dunnigan2001victory}: This deceives an attacker by blending a real object into a background or environment. For example, valuable files and programs can be given misleading names to make it difficult for an attacker to find them. 

\item {\em False information}~\cite{dunnigan2001victory}: Fake information (or disinformation) can be planted to mislead an attacker in cyberspace. However, most false information about cybersystems is easy to be checked by testing or trying it out. Therefore, this kind of deception cannot fool attackers for a long time.

\item {\em Lies}~\cite{dunnigan2001victory}: This provides deceptive information, which is false. However, lies are distinguished from disinformation because they are often provided as responses to questions or requests. In this sense, even if this deception uses false information, it implies a passive action because lies may not be used unless it is requested. Lies could be more effective than explicitly denying access as they encourage an attacker to continue wasting time trying to access the resource in different ways and at different times.

\item {\em Displays}~\cite{dunnigan2001victory}: This deception is  similar to {\em inventing} as it also aims at misleading objects or information. \new{However, unlike the inventing, which creates a new, fake object, `displays' can be applied to a real object (i.e., an object which is actually in a give system) by displaying it differently.}
 \end{itemize}
 
\subsubsection{Presence of Actual Objects or Information} Deception can be applied when an actual, true object exists or no actual object exists. To be more specific, a defender can use true objects or information but may want to hide it by lying or providing false information towards it. For example, even if a system uses Windows as an operating system but may lie it uses Unix. But the system is using a certain OS in deed. On the other hand, a fake object, which has not existed anywhere, can be created for the purpose of creating uncertainty or confusion, which can lead an opponent to reach a suboptimal or poor decision, such as honey tokens or honey files. 

\subsubsection{Expected Effects (or Intents) of Defensive Deception} Via defensive deception, one or more effects are expected as the process of achieving the goal of deception, making an attacker choose a suboptimal choice in its strategies. We categorize these expected effects in terms of the following five aspects: 
\begin{itemize}[leftmargin=*]
\item {\em Hiding}: This effect is addressed when deception is used to change the display of a real object into something else. Some obfuscation defense techniques also use hiding data to slow down the escalation of attackers or increase attack complexity~\cite{almeshekah2014planning, rowe2016introduction}.

\item {\em Luring}~\cite{almeshekah2016cyber, bell1991cheating}: This effect is shown when a fake object is demonstrated to an attacker where the object looks like a real object an attacker is interested in. The typical examples are honeypots to lure attackers or honey files to send false information to the attackers.

\item {\em Misleading}~\cite{daniel1982strategic}: Most deception techniques have this effect as the last step to achieve the ultimate goal of deception, misleading an attacker to choose a suboptimal action.

\item {\em Blending}: This effect is made when a real object can be well blended into background or environments. Its effectiveness relies on an environment and in altering the appearance
of things to hide. For instance, give password files a non-descriptive name to elude automated searches run by hackers looking for files named ‘pass' \cite{yuill2006using}. Additionally, blending hidden objects with the environment can be achieved through altering the environment instead. For example, a defender can create noise in the environment through adding bogus files to make it harder to find critical ones.

\item {\em Confusing}~\cite{almeshekah2014planning}: This effect comes from perceived uncertainty caused by a lack of evidence, conflicting evidence, or failing in discerning observations (e.g., cannot pinpoint whether a car on the street is red or blue)~\cite{almeshekah2014planning, Josang18}. A defender can confuse an attacker by presenting both truths and deceits~\cite{almeshekah2014planning}.
\end{itemize}
Some deception techniques use more than one technique, such as bait-based deception that includes lies to increase attack cost or complexity~\cite{keromytis2014systems}. \citet{bowen2009baiting} designed a trap-based defense technique to increase the likelihood of detecting an insider attack which combines several deception techniques like camouflage, false information, and creating fake objects. First, the deception system embeds a watermark in the binary format of the document file to detect when the decoy is loaded. Moreover, the content of each decoy document includes several types of ``bait'' information, such as online banking logins, login accounts for online servers, and web-based email accounts. A ``beacon'' is embedded in the decoy document that signals a remote web site upon opening of the document. 
    
\subsubsection{Ultimate Goals of Defensive Deception in Cyberspace}  The ultimate goal of a defender using deception techniques is either protecting assets or detecting attackers or both. Although some deception techniques allow the defender to achieve both asset protection and attack detection, deception by hiding is more likely to protect system assets while deception by using false objects or information is to catch attackers.

\subsubsection{Activeness of Defensive Deception}  \citet{caddell2004deception} categorized deception as \emph{passive} vs. \emph{active}. \emph{Passive deception} uses hiding the valuable assets or their capabilities and information from attackers. On the other hand, \emph{active deception} tends to use fake, false information aiming for the attacker to form false beliefs, leading to choosing suboptimal decisions. However, even if false information or lies are used, the defense goal can be either protecting assets or detecting attackers. Hence, the distinction of active deception and passive deception may not be always clear.

In Table~\ref{tab:technique-effect-goal}, we summarized how a particular deception technique can be understood based on the four key aspects: the presence of actual objects or information used in deception, expected effects (or intents), ultimate goal(s), and the activeness (or passiveness) of the used deception technique.

\section{Design Principles and Unique Properties of Defensive Deception} \label{sec:design-principles}
In this section, we discuss the four key design principles of defensive deception techniques. In addition, we address unique properties of defensive deception and their key merits and caveats when using defensive deception techniques. Further, we discuss how the defensive deception techniques differ from other similar defense techniques, such as moving target defense or obfuscation techniques.

\subsection{Design Principles of Defensive Deception}
In this section, we discuss the design principles of defensive deception in terms of the four aspects: {\em what-attacker-to-deceive} (i.e., what type of an attacker to deceive), {\em when-to-deceive} (i.e., when deception can be used), {\em how-to-deceive} (i.e., what particular deception technique can be used to deceive an attacker). We discuss each principle as follows:
\subsubsection{What-Attacker-to-Deceive} This design principle asks to determine what type of an attacker a defender wants to deceive. For example, if the defender targets to deceive attackers performing reconnaissance attacks as outside attackers, it may aim to deceive them by providing false information about a system configuration (e.g., saying using Windows operating system (OS) even if it actually uses Unix OS). In addition, if a valuable system asset should be protected from attackers aiming to exfiltrate confidential information to an outside network, the defender may deploy a honeypot which mimics a real node with highly confidential information (e.g., a database). Hence, determining {\em what-attacker-to-deceive} is to decide what attackers to target by the defender. Since developing a defensive deception technique incurs cost, there should be in-depth analysis and investigation on whether a given attacker should be prevented or detected by a defensive deception technique in terms of cost, effectiveness, and efficiency of its deployment.

\subsubsection{When-to-Deceive} This design principle refers to determining when a deception technique should be used in terms of the attack stage of a given attacker in the cyber kill chain (CKC)~\cite{chen2014study, okhravi2013survey}. The six CKC stages include reconnaissance, delivery, exploitation, command and control, lateral movement, and data exfiltration~\cite{okhravi2013survey}. Outside attackers mainly perform attacks in the stage of reconnaissance and delivery stages with the aim of penetrating into a target system. On the other hand, inside attackers aim to perform attacks to exfiltrate confidential information to the outside, unauthorized parties. For example, when more scanning attacks by the outside attackers are detected in the reconnaissance stage, the defender can use fake patches with false vulnerability information, which can lure the attacker to honeypots. After the attacker successfully got into the system, becoming the inside attacker, the defender can use honey files or tokens to lure the attacker which may think of exploiting them to perform attacks with the false information. \new{However, there are also legacy defense mechanisms, such as access control, intrusion detection and prevention, or emerging technologies, such as moving target defense as the alternative mechanisms to deal with the same types of attackers. In such cases, there should be utility analyses based on losses and gains in terms of the timing of using a defensive deception technique.}

\subsubsection{How-to-Deceive}  The design decision on how-to-deceive in employing defensive deception is related to what type of a defensive deception technique to use. The conceptual deception techniques have been discussed in Section~\ref{subsec:taxonomies-defensive-deception}, such as masking, mimicking, decoying, false information, baits, and so forth. However, what specific technique to use for defensive deception is more related to what technology to use to achieve deception. Some example defensive deception technologies include honeypots, honey files, honey tokens, fake patches, fake network topologies, fake keys, or bait files~\cite{han2018CSUR}.

\subsection{Benefits and Caveats of Defensive Deception Techniques} \label{subsec:benefits-caveats}
In this section, we discuss the key benefits of using defensive deception techniques as well as the caveats of developing and employing them. 

\subsubsection{Key Benefits of Defensive Deception Techniques}
\begin{itemize}[leftmargin=*]
\item Defensive deception is relatively cost-effective compared to other defense strategies because its deployment cost is relatively low while showing relatively high effectiveness to mislead attackers. For example, honey files or honey tokens are relatively simple to deploy and maintain, compared to other traditional cybersecurity mechanisms, such as access control or intrusion detection.

\item Defensive deception provides complementary defense services to other legacy defense mechanisms, such as intrusion detection or prevention. For example, honeypots are well known as an effective monitoring mechanism that can provide additional attack features to enhance intrusion detection as well as proactively to protect a system from intrusions before they actually launch attacks to targets.

\item Various types of defensive deception techniques are deployable at multiple layers of systems, such as network, system, application, and data layers~\cite{han2018CSUR}, without significant changes of existing system architectures. High deployability of defensive deception techniques also provides design flexibility as well as defense-in-depth (DiD) capability with multiple layers of protections. 

\item Automated cyberdeception techniques, such as obfuscating session information, data flow, and software's code obfuscation, have significantly improved security to counter automated attacks.

\end{itemize}

\subsubsection{Key Caveats of Defensive Deception Techniques}
\begin{itemize}[leftmargin=*]
\item Since it is highly challenging to obtain an attacker's motivation, intent, and goal in launching and executing a particular attack, it is not trivial to choose an optimal defensive deception strategy based on the estimated risk and benefit associated with a given defensive deception strategy.
\item Most honey-X techniques (e.g., honeypots, honey files, honey tokens) aim to mislead attackers to choose suboptimal or poor choices in launching attacks by false information. This may introduce extra procedures or protocols for normal users or a defender not to be confused by them.

\item Due to the nature of deception, the effectiveness of defensive deception requires continual configuration, reconfiguration, and implementation process. Otherwise, adversaries would be able to easily distinguish deceptive devices unless deception is automated like software obfuscating techniques.

\item With the emergence of clever attackers, there is a growing demand for high-interactive honeypots or elaborated deception strategies that cannot be easily identified by attackers. This brings a higher defense cost and more difficulty in the management for a defense system than a traditional system without defensive deception mechanisms.
\end{itemize}

\subsection{Distinctions between Defensive Deception and Other Similar Defense Techniques}

Defensive deception is often mentioned as moving target defense or vice versa. In addition, obfuscation has been used to achieve the same aim like defensive deception. In this section, we discuss how the roles and natures of these three techniques differ from each other as well how they overlap in their functionalities and aims.

\subsubsection{Distinction with Moving Target Defense (MTD)}  MTD is similar to defensive deception in terms of its aim to increase confusion or uncertainty of attackers, leading to deterring the escalation of their attacks to a next level or failing their attacks. However, the key distinction is that MTD does not use any false information to actively mislead attackers while defensive deception often involves using false objects or information for the attackers to form false beliefs and be misled to make suboptimal or poor attack decisions. MTD relates its key functionality mainly with how to change system configurations more effectively and efficiently while defensive deception more involves the exploitation of manipulating the attacker's perception. In this sense, if defensive deception is well deployed based on a solid formulation of manipulating the attacker's perception, it can be more cost effective than deploying MTD. \citet{ward2018MIT} considered MTD as part of defensive deception while \citet{Cho20-mtd} treated defensive deception as part of MTD. The distinction between these two is not crystal clear because defensive deception can be deployed using MTD techniques, such as randomness or dynamic changes of system configurations (e.g., dynamically assigning decoys in a network) while it may not use false information all the time (e.g., baits or omission such as no response).
 
\subsubsection{Distinction with Obfuscation} Obfuscation techniques have been used by an attacker to obfuscate malware~\cite{You10-Obfuscation}, metamorphic viruses~\cite{Borello08}, or malicious JavaScript code~\cite{Xu12-obfuscation}. In terms of a defender's perspective, obfuscation techniques have been used to obfuscate private information~\cite{Ardagna07-location-obfuscation}, or Java bytecode for decompiling~\cite{Chan04-jss}. Although obfuscation techniques are often mentioned as the part of defensive deception techniques, they have been studied as a separate research area for decades. \citet{Chan04-jss} conducted an extensive survey on obfuscation techniques enhancing system security and discussed the key aims of obfuscation techniques as: (i) increasing the difficulty of reverse engineering of the program logic; (ii) mitigating the exploitability of vulnerabilities by attackers; (iii) preventing modifications by unauthorized parties; and (iv) hiding data or information. Although the concept of defensive deception has gained popularity of late while obfuscation techniques have been applied for decades, it is quite obvious that the aim of defensive obfuscation is well aligned with that of defensive deception. However, a subtle difference lies in that defensive deception involves cognitive aspects of an attacker's perception while defensive obfuscation is directly involved with achieving security goals (e.g., confidentiality, data integrity, availability). Defensive obfuscation techniques are also used in combination with diversification~\cite{Chan04-jss} (e.g., diversifying systems, network configurations, or components), which is a well-known concept used by MTD techniques.

\begin{figure}
    \centering
    \includegraphics[width=0.5\textwidth]{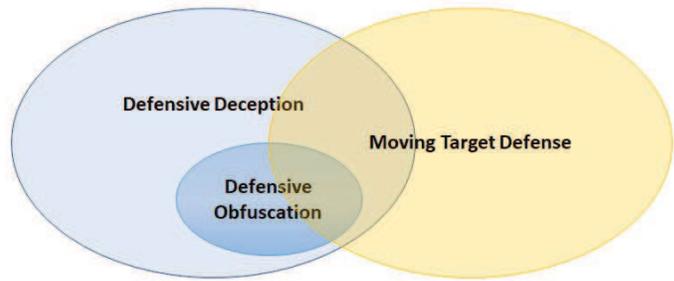}
    \caption{Relationships between defensive deception, moving target defense, and defensive obfuscation.}
    \label{fig:relationships-dd-mtd-do}
    \vspace{-2mm}
\end{figure}
Based on our understanding on the functionalities and aims of defensive deception, MTD, and defensive obfuscation, we would like to hold our view based on Fig.~\ref{fig:relationships-dd-mtd-do}, which shows how each domain is related to each other.  Basically, our view is that a defensive obfuscation technique can belong to either MTD or defensive deception. In addition, there are also an overlapping area of MTD and defensive deception where some defensive techniques require using the features of both concepts.


\section{Game-Theoretic Defensive Deception (GTDD) Techniques}
\label{sec:game-dd}


In this section, we discuss the key component of modeling defensive deception techniques using game theory. In addition, we discussed a wide range of defensive deception techniques that have been considered based on various types of game theory along with their pros and cons.

\subsection{Key Components of Game-Theoretic Defensive Deception}
\label{subsec:key-components-game}

Game theory has been used extensively in modeling cybersecurity problems as a framework to model the following key elements of cybersecurity problems: a defender and attacker as players, actions as attack and defense strategies, observability of a system and an opponent's strategies, or system dynamics. We discuss these key elements of cybersecurity games as follows.

\subsubsection{Players} Most research taking game-theoretic defensive deception approaches model a two-player game where the two players are an attacker and a defender using defensive deception~\cite{kiekintveld2015game, mao2019game, pawlick2015deception}. However, even in a two-player game, some deception games modeled different types of players under each player. \citet{pawlick2015deception} modeled a two-player signaling game where a defender as a sender can be either a normal system or a honeypot while an attacker as a receiver has one type. Depending on the type, the defender takes its strategy.  \citet{pawlick2015flip} introduced a three-player game between a cloud defender, an attacker and a device which is connected with the cloud where the defender can send false signals to deceive the attacker.

\subsubsection{(In) Complete Information}  A complete information game allows players to have full knowledge of the game parameters, such as possible actions to be taken by other players, a reward function, and the current state of the game in case of dynamic or multistage games. A game with complete information is sometimes considered to be impractical. The assumption of the defender knowing the set of attack actions to be taken by its adversary is not realistic due to inherent uncertainty in a given context. On the other hand, an incomplete information game represents a practical class of games where one or more players may or may not know some information about the other players~\cite{kajii1997robustness}. For instance, a player may not fully know other players' types, strategies, and payoff functions. 

\subsubsection{(Im) Perfect Information}  Imperfect information game represents a game in which a player may not know what exact actions have been played by other players in a given game. This makes it computationally prohibitive to track the history of actions in case of a multistage game. However, all players may know their opponents' types, strategies, and payoff functions, which can form a complete, imperfect information game.  This imperfect information game is often assumed in cybersecurity games between an attacker and a defender~\cite{basar1999dynamic, phlips1988economics}. 

\subsubsection{Partial Observability}  Players acting upon dynamic systems are modeled as dynamic and stochastic games. In this case, the system/environment state changes over time and depends on the actions taken by all the players. However, in some scenarios, one or more players will not fully observe the state. This results in a partially observable game. If the system environment is affected by an exogenous factor, the state transition is stochastic which results in a partially observable stochastic game (POSG)~\cite{hansen2004dynamic}. In the special case of POSG with only a single player, the game turns to a partially observable Markov dynamic process~\cite{cassandra1998survey}. POSG model deals with the most general form of games that capture different game settings, such as incomplete information as well as imperfect information (i.e., imperfect monitoring). Therefore, solving such a game efficiently is still an open research question.

\subsubsection{Bounded Rationality} Both the rational and bounded rational players are commonly applied in a game between an attacker and a defender.
A rational player can always choose an optimal strategy to maximize its expected utility. However, a bounded rational player has only limited resources and cannot afford an unlimited search to find an optimal action~\cite{Tadelis13}. 

\subsubsection{Pure or Mixed Strategies}  Two main strategies in game theory are pure strategy or mixed strategy~\cite{Tadelis13}. A pure strategy means a player's strategy is determined with the probability 0 or 1. In contrast, a mixed strategy is a probability distribution of several pure strategies.  For example, \citet{clark2012deceptive} used pure strategies to perform deceptive routing paths to defend against jamming attacks in a two-stage game using Stackelberg game theory. \citet{zhu2012deceptive} also used deceptive routing paths against jamming attacks but identified an optimal routing considering resource allocation based on mixed strategies.

\subsubsection{Uncertain Future Reward in Utility}  Due to the inherent uncertainty caused by multiple variables that may control the future states, a player's utility cannot be guaranteed~\cite{Tadelis13}. In game-theoretic defensive deception research, uncertainty is mainly caused by non-stationary environmental conditions and non-deterministic movements (or actions) by opponent players. In a signaling game, uncertainty is caused by both the movement of an opponent player and its type~\cite{pawlick2015deception,mohammadi2016Springer} where a player can take an action after it knows its opponent's type.

\subsubsection{Utilities} For a player to take an optimal action, it is critical to know both its own payoff (or utility or reward) function and its opponent's payoff function. In a static, one-time game, a reward per strategy is based on action profiles (i.e., a set of actions each player can take) of all players~\cite{Tadelis13}. However, in a sequential game (i.e., a repeated game), a reward can be estimated based on the history of action profiles from the beginning to the end of the game. Hence, in such games, a player aims to maximize the expected average or accumulated reward over the period of the sequential game~\cite{basar1999dynamic}. 

\subsubsection{Full Game or Subgame} A full game represents a game where each player considers a set of all possible actions to select one action to maximize its utility. A subgame indicates a subset of the full game where a player considers a subset of all possible actions when selecting one action to maximize its utility. Particularly, in sequential games, a subgame includes a single node and all its successors~\cite{Tadelis13}. It is similar to a subtree in a tree-structure sequential game, which refers to an extensive form game. \citet{House10} used the hypergame theory to model the interactions between an attacker and a defender. Hypergame uses a subgame to model each player's different view on the game and corresponding strategies under the subgame. In addition, \citet{Zhang19-subgame} and \citet{xu2016modeling} used {\em Subgame Perfect Nash Equilibrium} (SPNE)~\cite{mas1995microeconomic} (i.e., a Nash equilibrium in every subgame) to deal with the resource allocation for both the defender and attacker in non-hypergames.

\subsubsection{\new{Static Game vs. Dynamic Game}} \new{In game theory, a static game is considered as one-time interaction game where players move simultaneously. Hence, a static game is a game of imperfect information. On the other hand, a dynamic game is considered as a sequential game assuming that players interact through multiple rounds of interactions as repated games. Although most dynamic games are considered sequential games with perfect information, if a game considers partial (or imperfect) observability, the dynamic game can be sequential across rounds of games but within a subgame, it can be a game of imperfect information (i.e., players know other players' moves in previous rounds but do not know their moves in a current round).}

\new{Game theory has provided a powerful mathematical tool for solving strategic decision making in various dynamic settings. Since the powerful capability of game theory has been clearly proven based on the achievement of a number of game theorists as Nobel Prize Laureates in the past~\cite{Tadelis13}, there is no doubt that game theory has substantially inspired in solving highly complicated, dynamic decision-making problems with mathematically validated solutions.  Although game theory has been substantially explored to provide solutions with solid theoretical proofs, one of its common limitations has been discussed in terms of too strong assumptions to derive Nash Equilibria, such as common knowledge about probability distributions of Nature's moves or players' correct beliefs towards opponents' moves. As game theory has a long history since 1700s~\cite{bellhouse2007problem}, its long evolution has made significant advancement in considering more and more realistic aspects of real-world problems. The exemplary efforts in addressing realistic scenarios include the consideration of bounded rationality, decision making under uncertainty using the concept of mixed strategies, games of imperfect or incomplete information, hypergame with different subjective views of players, games with partially observable Markov Decision Process (POMDP), subjective beliefs, subjective rationality, and so forth~\cite{Tadelis13}. We believe these efforts have obviously opened a door for game theory to be highly applicable in real systems, including systems concerning cybersecurity. In the context of developing defensive deception techniques, game theory can contribute to manipulating an attacker's beliefs and accordingly leading the attacker to miscalculate utilities under uncertainty with the aim of failing the attacker.  In addition, how frequently a certain defense should be executed or what defense strategy to choose can be also determined based on game theoretic decision methods.}

\subsection{\new{Common Games Formulated for Defensive Deception}}

\new{We now discuss popular game theories used to formulate various defensive deception techniques. We will discuss the following games:
\begin{itemize}
\item {\em Bayesian games}: A Bayesian game is a game of incomplete information which means part of or all players do not know their opponents' types, actions, or payoffs where each player knows their own type, a set of actions, and corresponding payoffs. Hence, each player has a subjective prior probability distribution of their opponent's type~\cite{harsanyi1967games}.
\item {\em Stackelberg games}: Stackelberg game refers to a sequential two-player game where the first player is a leader and the second player is a follower. In a stage game, the leader moves first, then the followers observe the action of leader and select their strategy based on the observation~\cite{von2010market}. Hence, the first player takes the first-mover-advantage which can lead the second player's move to their intended direction~\cite{Tadelis13}.
\item {\em Signaling games}: This is a sequential Bayesian game where one player sends a signal (i.e., information) to another player. When the player sends false information, it can be most costly~\cite{Tadelis13}.
\item {\em Stochastic games}: This is a multistage game in which the game evolves from one stage (state) to another following a stochastic distribution. The transition probability matrix depends on a set of action profiles of both players and may also depends on random exogenous factors of the game environment. The game dynamics determine the payoffs of both players~\cite{shapley1953stochastic}.
\item {\em Games with Partially Observable Markov Decision Process (POMDP)}: The POMDP considers to model an agent's decision process where the agent selects actions to maximize a reward of the system. However, the agent cannot directly perceive the system state, but it can obtain observations depending on the state. The agent can maintain a belief of a system state based on the observations.  Formally, a POMDP model can be represented by a tuple $(S,A,\Omega,T,R,O )$~\cite{cassandra1998survey}, where $S$ represents a set of states, $A$ represents a set of actions, $\Omega$ represents a set of observations, $T$ is a state transition function, $R$ is a reward function, and $O$ is a set of conditional observation probabilities. As an extended version of the POMDP, the Interactive-POMDP (IPOMDP) presents a multiagent setting which allows an agent to model and predict behavior of other agents~\cite{gmytrasiewicz2005framework}. A Bayes-Adaptive POMDP (BA-POMDP) assumes that transition and observation probabilities are unknown or partially known~\cite{ross2007bayes}.
\end{itemize}
}

\subsection{Game-Theoretic Defensive Deception} \label{subsubsec:GTDD}
In this section, we discuss game-theoretic defensive deception techniques used in the literature. We use the following classification to discuss game-theoretic defensive deception techniques for asset protection: honeypots, honeywebs, honeynets, obfuscation, deceptive signals, \new{multiple defensive deception techniques}, fake objects, honey patches, deceptive network flow. Some studies used empirical game-theoretic experiment settings where players are human subjects. We also discussed them to show how game theory has been used to develop defensive deception in empirical experimental settings with human-in-the-loop.

\subsubsection{Honeypots} A honeypot has been studied as the most common defensive deception strategy in the literature. A honeypot is mainly used in two forms: {\em low interaction honeypots} (LHs) and {\em high interaction honeypots} (HHs)~\cite{han2018CSUR}.  The differences between LHs and HHs are based on different deception detectability and deployment cost. HHs provide lower detectability by attackers than LHs; However, HHs incurs higher deployment cost than LHs~\cite{han2018CSUR}.

\citet{pawlick2015deception} employed a signaling game to develop a honeypot-based defense system where an attacker has the ability to detect honeypots. The authors investigated multiple models of signaling games with or without evidence when complete information is available or not. In the cheap-talk games with evidence (i.e., a signaling game with deception detection), extended from cheap-talk games (i.e., costless communication signaling game between a sender and receiver), the receiver (i.e., an attacker) is modeled to detect deception (i.e., a honeypot) with some probability. They found that the attacker's ability to detect deception does not necessarily lower down the defender's utility. 

The signaling game with evidence has been also applied to model the honeypot selection or creation in other works~\cite{pawlick2018modeling, pibil2012game}. \citet{pawlick2018modeling} used a signaling game with evidence where the evidence is estimated based on a sender type and a transmitted message. The authors extended the signaling game with a detector with probabilistic evidence of deception. However, this signal is vulnerable to attackers which can analyze the signal, leading to detecting the deception and causing the evidence leakage. \citet{pibil2012game} introduced a game-theoretic high-interaction honeypot to waste attackers' resources and efforts. The authors designed a {\em honeypot selection game (HSG)} as a two-player zero-sum extensive-form game with imperfect and incomplete information. A honeypot is designed to mimic servers with high importance to provide cost-effective defensive deception.

\citet{la2016deceptive} proposed a two-player attacker-defender deception-based Bayesian game in an IoT network. They used honeypots as a deceptive defense mechanism. They studied the Bayesian equilibrium in one shot and repeated games. Their findings showed the existence of a certain frequency of attacks beyond which both players will primarily take deception as their strategy. The authors used a honeypot as a deceptive strategy where a game is considered under incomplete information.

\citet{cceker2016deception} proposed a deception-based defense mechanism to mitigate Denial-of-Service (DoS) attacks. The defender deploys honeypots to attract the attacker and retrieve information about the attacker's real intent. They used signaling games with perfect Bayesian equilibrium to model the interactions between the defender and attacker. \new{\citet{basak2019identifying} used cyber deception tools to identify an attacker type as early as possible to take better defensive strategies. The attacker's type is reflected in actions and goals when planning an attack campaign. The authors leveraged on a multistage Stackelberg Security game to model the interaction between the attacker and defender. In the game, the defender is a leader taking strategies considering the attacker’s strategy. As a follower, the attacker selects its strategy after observing the leader's strategy. Through a game-theoretic approach, the defender selects deception actions (e.g., honeypots) to detect specific attackers as early as possible in an attack.} 

\citet{mao2019game} used honeypots as a defense strategy in a non-cooperative Bayesian game with imperfect, incomplete information where an attacker is a leader and a defender is a follower. The authors considered a player's perception towards possible motivation, deceptions, and payoffs of the game. \new{\citet{kiekintveld2015game} discussed several game models that address strategies to deploy honeypots, including a basic honeypot selection game. They first developed a honeypot selection game as a two-player zero-sum extensive-form game with imperfect and incomplete information in~\cite{pibil2012game}. Then they extended the game to allow additional probing actions by the attacker in which attack strategies are represented based on attack graphs~\cite{kulkarni2020deceptive, xi2020hypergame}.} The authors conducted experimental performance analysis to compare the performance of these model and discussed the advantages and disadvantages of the considered game-theoretic models in the defensive deception research.

\citet{durkota2015optimal} leveraged a Stackelberg game where an attacker follows an attack graph and a defender can deploy a honeypot to deceive the attacker based on an efficient optimal strategy searching method using Markov Decision Processing (MDP) and sibling-class pruning.  \citet{Kulkarni20,Stephanie2020Harnessing,Olivier2020Partially} investigated game-theoretic cyberdeception techniques played on attack graphs. 
\new{\citet{Kulkarni20} developed a zero-sum, hypergame of incomplete information for evaluating the effectiveness of the proposed honeypot allocation technique.}
\citet{Stephanie2020Harnessing} developed a Stackelberg game to allocated defensive resources and manipulate a generated attack graph. \citet{Olivier2020Partially} studied the malicious behavior through an epidemic model and solved a one-sided partially observable stochastic game for cyberdeception where a defender distributes a limited number of honeypots. 

\new{\citet{wagener2009self} addressed the issue of a honeypot being overly restrictive or overly tolerant by proposing a self-adaptive, game theoretic honeypot. The authors modeled the game between an attacker and a defender by reusing the definitions proposed in~\cite{greenwald2007matrix}. They applied game theory to compute the optimal strategy profiles based on the computation of Nash Equilibrium.} The game theory directs the configuration and a reciprocal action of the high-interaction honeypot. In their experiment, this technique is capable of deriving an optimal strategy based on Nash equilibrium with rational attackers.  \citet{aggarwal2016cyber, aggarwal2017modeling} discussed a sequential, incomplete information game to model the attackers' decision-making in the presence of a honeypot and combined this model with instance-based learning (IBL). The authors introduced two timing-based deception by applying early or late deception in the rounds of games and investigated the effect of timing and the extent of deception.  Their results showed that high amount of using deception and late timing of deception can effectively decrease attack actions on the network.  However, using a different timing and amount of deception \new{did not} introduce any difference in attacking on a honeypot action.

Various honeypot allocation methods for deception over attack graphs have been studied based on game theoretic approaches that considers uncertainty using partially observable MDP (POMDP) and stochastic games (POSG)~\cite{anwar2020gamesec,kamhoua2018game}.  \citet{anwar2019game} developed a static game model to protect a connected graph of targeted nodes against an attacker through honeypot placements based on node features and significance. The game model is extended to study the game dynamics as a stochastic allocation game in~\cite{anwar2020Honeypotbook}. To resolve high complexity of attack graph-based honeypot allocation games, the authors proposed a heuristic approach.  

\new{\citet{nan2019behavioral} and \citet{nan2020development} provided Nash equilibrium-based solutions for a defender to intelligently select the nodes that should be used for performing a computation task while deceiving the attacker into expending resources for attacking fake nodes.  \citet{nan2019behavioral} investigated a two-player static game of complete information with mixed-strategy, where the system selects a node to place the deception source and the attacker selects a node to compromise.  \citet{nan2020development} constructed a static game with mixed-strategy, where both players select one strategy (target node) with probability distribution. The defender selects a node to install defense software while the attacker selects a target node to compromise and avoid being detected by the anti-malware software.}

\vspace{1mm}
\noindent {\bf Pros and Cons}: A honeypot is the most popular defensive deception technology which has matured over decades. Since its aim is not to introduce any additional vulnerabilities, if an attacker is successfully lured by the honeypots, the honeypots can protect the existing system components (i.e., system assets) while collecting additional attack intelligence which can lead to improving intrusion detection with new attack signatures. However, maintaining honeypots incurs extra cost. In addition, there is still a potential performance degradation due to the existence of the honeypots, such as additional routing paths or resource consumption. However, the drawbacks of honeypots have been little investigated. In addition, as more intelligent, sophisticated attackers have been emerged recently, it becomes more challenging to develop realistic honeypots for effectively deceiving attackers in terms of both complexity and cost.

\subsubsection{Honeywebs}
\new{\citet{el2018new} proposed a web server framework that provides a web application firewall with a honeyweb to detect malicious traffics and forward suspicious traffics to honeypot servers. This work formulated the interactions between the attacker (i.e., malicious traffics) and the defender (i.e., honey servers) using a game of imperfect information (i.e., players do not know their opponent's moves) and complete information (i.e., players know their opponent's type, action, and payoff function) because they consider simultaneous moves by both the attacker and defender.}  Typically, this research combines deception technologies, including honey token, honeypot, and decoy document. In the game definition, the main action of a defender is implementing a virtual machine (VM) to consume the attacker's effort and resources. 

\vspace{1mm}
\noindent {\bf Pros and Cons}: Honeywebs are rarely used particularly based on game-theoretic approaches. They are used to assist honeypots along with honey tokens or honey files. Developing fake webpages incurs less cost while its role is more like an assistant to support the honeypots. Hence, honeywebs may not be used without other defensive technologies.

\subsubsection{Honeynets} \citet{garg2007deception} considered an attacker and a honeynet system (i.e., a defender) as players in a strategic non-cooperative game. The framework is based on extensive games of imperfect information. In their game model, the defender makes deception about the placement of a honeypot while the attacker can probe the target host to identify their true roles with some probability. The attacker's utility considers the cost of both probing and compromising a host or honeypot. They studied the mixed strategy equilibrium solutions of these games and showed how they are used to determine the strategies of the honeynet system. 

\new{\citet{dimitriadis2007improving} proposed a honeynet architecture based on an attacker-defender game. This architecture is called 3GHNET (3G-based Honeynets) and aimed to enhance the security of the 3G core network by defending against DDoS and node compromise attack. The main components of this 3GHNET are two gateways with a set of strategies to control and capture the data flow between two nodes. In their emulation, each gateway and attacker are considered as individual players. The 3GHNET-G is a two-player, non-cooperative and zero-sum game. Leveraging on Nash equilibrium, the authors calculated the payoff matrix to assist the players to search for the optimal strategy.} 

\vspace{1mm}
\noindent {\bf Pros and Cons}: The honeynet is a system architecture and mainly designed to deal with inside attackers and allow a system manager to monitor threats and learn from them. The Honeynet Project~\cite{spitzner2003honeynet} is a typical example of a honeynet consisting of multiple honeypots applied in practice. Since the honeynet works with multiple honeypots, how to optimally deploy and work together among them needs more investigation based on both the effectiveness of deception and additional deployment costs.

\subsubsection{Obfuscation} \citet{Shokri15} modeled a leader-follower game (i.e., Stackelberg game) between a designer of an obfuscation technique and a potential attacker and designed adaptive mechanisms to defend against optimal inference attacks. They assumed that the users plan to protect sensitive information when sharing their data with untrustful entities. This design allows the users to obfuscate the data with noises before sharing it. The attacker can observe valuable information of users and noises caused by obfuscation. They provided linear program solutions to search for an optimization problem that provably achieves a minimum utility loss under those privacy bounds. 

\citet{hespanha2000deception} proposed a defensive deception framework based on non-cooperative, zero-sum, stochastic games with partial information. They analyzed how a defender in a competitive game can manipulate information available to its opponents to maximize the effectiveness of the defensive deception. They found that there exists an optimal amount of information to be presented to effectively deceive an attacker.

\vspace{1mm}
\noindent {\bf Pros and Cons}:  Compared to other defensive deception techniques, such as honey-X techniques, the key benefit of data obfuscation is easy deployability with low cost. However, adding noise into normal information can also confuse a defender or a legitimate user. On the other hand, most data obfuscation research mainly aims to develop a technique of how to hide real information rather than how to detect an attacker.

\subsubsection{Deceptive Signals} In many game-theoretic approaches that consider defensive deception as a defense strategy, deception is simply used as a signal to deceive an opponent player. \citet{pawlick2015flip} modeled a game of three players, consisting of a cloud defender, an attacker and a device, in a cloud-based system that can deal with advanced persistent threats (APTs). The authors designed the so-called {\em FlipIt} game to model the interactions between an attacker and a defender where signaling games are used to model the interactions between the device and the cloud which may be compromised with a certain probability.  \citet{xu2016modeling} designed a game between an attacker and a defender where the attacker has the ability to investigate the vulnerability of a target and the defender can apply defensive technologies to change the vulnerability of a certain device. This work mainly studied how the attackers' preparation for launching an attack affects the effectiveness of defensive deception strategies. As the result, the authors applied the subgame perfect Nash Equilibrium (SPNE) to analyze the strategic interactions of the terrorist's costly learning and the defender's counter-learning.

\citet{yin2013optimal} leveraged a fake resource or converted a real resource to secrete recourse to mislead attackers. They applied a Stackelberg game to model the interactions between an attacker and a defender. The authors conducted the mathematical analysis of a game using both pure and mixed strategies played by the defender. They provided an optimal strategy for the defender and calculated the accuracy rate of identifying when the attacker can deploy unlimited surveillance before attacking. \citet{horak2017manipulating} used one-sided POSG (partially observable stochastic games), in which a defender has perfect information while other players have imperfect information.  In this work, an attacker and defender take sequences of actions to either deceive an opponent or attempt to obtain true actions taken by the opponent. This work assumed that a rational attacker has knowledge towards its opponent and its actions to take. However, the attacker has no knowledge of a network topology which introduces a hurdle to identify a target and whether the defender is aware of its presence in the network. 

\citet{ferguson2019game} proposed a hypergame-based defensive deception scenario where players have imperfect and incomplete information. Based on the key concept of a hypergame, the authors modeled an individual player's own game where the game structure and payoffs may be manipulated by an opponent player. However, this work assumed an asymmetric view by the attacker and defender in that the attacker does not know the defenders taking defensive deception strategies while the defender can know the attacker's true payoff and game structure. 

\citet{bilinski2019you} introduced game-theoretic deception procedures based on a masking game in which a defender masks the true nature of a device. In this game, an attacker can ask the defender whether a given device is real or fake at each round of the game. The defender needs to pay cost if it lies.  After several rounds, the attacker needs to choose a target to attack. The authors investigated this scenario by conducting a mathematical analysis for non-adaptive and adaptive game models.  They also designed a Stackelberg game model where the attacker is a leader and the defender is a follower.  Through an Markov Decision Process (MDP) simulation, they investigated the potential behavior of an attacker at noncritical points.

\vspace{1mm}
\noindent {\bf Pros and Cons}: This deceptive signal is an abstract way to apply a game theory concept to model a deception game between an attacker and a defender. Due to the nature of the abstracted game formulation, it has a high flexibility that can use various types of deception techniques. On the other hand, it is quite challenging to model a cybersecurity problem in a particular context as the proposed deception game framework does not use specific defensive deception techniques. 

\subsubsection{Multiple Defensive Deception Techniques} Some existing approaches leverage a set of cyberdeception technologies to model a set of strategies by a defender. 
\citet{chiang2018defensive} discussed defensive deception to improve system security and dependability based on an SDN environment by utilizing a set of honey technologies. The authors discussed what are the critical requirements to realize effective defensive deception and identified promising evaluation methods including metrics and evaluation testbeds.  \new{\citet{chiang2018defensive} constructed a dynamic game of complete information where the attacker does not know the strategies chosen by defender.}  \citet{huang2019dynamic} discussed several defensive deception techniques, such as honeypots, fake personal profiles, or multiple game models. They used a static Bayesian game to capture stealthy and deceptive characteristics of an attacker and advance this model by applying asymmetric information one-shot game. They demonstrated the performance of their proposed techniques under APT based on the Tennessee Eastman (TE) process as their case study.

\noindent {\bf Pros and Cons}: Since multiple defensive deceptions are combined and used as a set of a defender's strategies, the defender can make more choices to defend against attackers based on different merits that can introduce to different types of attacks. In particular, to deal with APT attackers which can perform multi-staged attacks, using various types of defensive deception can provide more relevant resources to deal with them. However, combing more than one deception techniques introduces additional deployment costs.  In addition, it is not trivial to identify a optimal combination of multiple defensive deception techniques.

\subsubsection{Fake Objects} \citet{mohammadi2016Springer} analyzed two-player signaling games with a defensive fake avatar for selecting an optimal strategy under various scenarios.  A defender can use an avatar in the signaling game as a fake internal user to identify an external attacker by interacting with external users. This game considered the payoff of two players and derived an optimal threshold of raising alarms. In their experiment, the expected payoff is used a metric to guide actions of the avatar or the defender.  Unlike other works using signaling games~\cite{pawlick2015deception, pawlick2015flip}, their signaling games put the defender as a second mover (i.e., a receiver) while an attacker is a first mover (i.e., a sender) where the games are played with incomplete information.  The key idea of using the signaling games is creating uncertainty for the attacker because the defender uses a fake identity which can make the attacker doubtful about whether or not the receiver is a real user. 

\citet{casey2015compliance} discussed interactions between an insider attacker and a defender of an organization. They considered a hostile agent that may obtain organization surface and harm the whole system. In order to mitigate this kind of threats, the authors proposed a \emph{honey surface} to confuse the malicious agent by designing a basic compliance signaling game to model the interaction between agents and the organization.  \citet{casey2016compliance} further discussed the insider threat in an organization and applied a signaling game to model the interactions between agents with learning intelligence and a defender.

\citet{thakoor2019cyber} introduced a general-sum game, named {\em Cyber Camouflage Games} (CCGs), to model the interactions between a defender and an attacker performing reconnaissance attacks. The defender can mask the machine in the network with fake information, such as an operating system, to mitigate the effect of reconnaissance attacks. To identify an optimal strategy, they introduced the Fully Polynomial Time Approximation Scheme (FPTAS) given constraints and applied Mixed Integer Linear Program (MILP) to search for an optimal strategy.

\citet{Zhu14-sh} simulated the infiltration of social honeybots-based defense into botnets of social networks. The authors proposed a framework, so-called SODEXO (SOcial network Deception and EXploitation) consisting of Honeybot Deployment (HD), Honeybot Exploitation (HE), and Protection and Alert System (PAS). HD is composed of a moderate number of honeybots. The HE considered the dynamics and utility optimization of honeybots and botmaster by a Stackelberg game model. The PAS chose an optimal deployment strategy based on the information gathered by honeybots. The results showed that a small number of honeybots can significantly decrease the infected population (i.e., a botnet) in a large social network.

\vspace{1mm}
\noindent {\bf Pros and Cons}: Although using fake identities or avatars as a defensive deception technique is less costly and relatively simple, it can introduce confusion to normal users and accordingly increase false alarms. However, little work has investigated the adverse effect of using fake identities, such as introducing confusion for legitimate users or producing extra vulnerabilities.  In addition, the effect of intelligent attackers with high deception detectability has not been studied.

\subsubsection{Honey Patches}
A honey patch mainly has two components: (1) A traditional patch to fix known software vulnerabilities; and (2) additional code to mislead an attacker to fake software vulnerabilities~\cite{Avery17}. In the literature, fake patches are also called \textit{honey patches}~\cite{Araujo14} or \textit{ghost patches}~\cite{Avery17}. 

\new{\citet{Araujo14} coined the term {\em honey patches} that can function as same as a regular patch; but it has the ability to efficiently redirect an attacker to a decoy, and allow the attacker to achieve a fake success. The game used in~\cite{Araujo14} is a dynamic game of incomplete information where the payoff of attacker is unknown to defender.}  In addition, the authors provided a strategy allowing a web server administrator to transfer regular patches into honey patches.  Enterprise scale systems are often exposed by the vulnerabilities due to the lack of boundary checking and subtle program misbehavior~\cite{Avery17}. A common way to eliminate the vulnerabilities is releasing security patches. However, as the security patches contain the location and types of vulnerability to be patched, an attacker could obtain a blueprint of system vulnerabilities by analyzing the security patches from the past. As a solution, \citet{Avery17} provided a fake patch technique for misleading the attackers that try to analyze the historical patches. In particularly, this fake patch technique is designed to protect the patch that fixes input validation vulnerabilities. This fake patch can contain two parts. One is creating a decoy patch file to seduce the attacker and attract the attacker's attention away from the real patch file and the other is to alert defenders of potential intrusions or exfiltration attempts. This decoy file is generated by a modified technique~\cite{bowen2009baiting}.  Another part is a bogus control flow for programs. This alternation misleads the reverse engineering, but does not change the output of the program. In their experiment, this technique was evaluated in~\cite{cadar2008klee} where the program runtime and program analysis measure the effectiveness of this technique. \new{The game and its components in~\cite{Avery17} were not rigorously presented since it mainly focused on the technicalities of developing fake honeypots. However, the deception and the evaluation of its reward represents a normal form one-shot game. Such games were clearly presented in~\cite{avery2018CS} where the authors evaluated the effectiveness of three deceptive patches (i.e., faux, obfuscated, and active response) and applying them into the proposed game-based module which provides the security guidelines.} In addition, they found that many techniques are unable to meet the proposed security definition while emphasizing the importance of assessing deceptive patches based on a clear and meaningful security definition.

\citet{Cho19-hgt} modeled a deception game based on hypergame theory in which an attacker and a defender have different perceptions in a given game. This work examined how a player's (mis)perception can affect its decision making to choose strategies to take, which can affect the player's utility in the given game. This work used Stochastic Perti Nets to build a probabilistic model of the hypergame where the defender uses fake patches as a deception strategy for misleading the attacker to believing in the fake patches and choose a non-vulnerable node as a target to compromise, which can lead the attack failure. 

\vspace{1mm}
\noindent {\bf Pros and Cons}: 
Honey patches are known as effective deception techniques to deceive attackers with low cost. However, as discussed in~\cite{avery2018CS}, when some systematic and meaningful security assessments are conducted to validate the security of the honey patches, some honey patches fail to pass those security criteria, such as indistinguishablity between real and fake patches, vulnerabilities of fake patches, or cryptographic breakability for obfuscated messages~\cite{avery2018CS}. 

\subsubsection{Deceptive Network Flow}

\citet{clark2012deceptive} proposed the deceptive network flow, representing the flow of randomly generated dummy packets. They assumed that both of real and deceptive packets are encrypted.  Hence, an attacker cannot distinguish between them and may spend limited resources on targeting a false flow. This work used this deceptive flow to lure the attacker and waste its resource to assist those real packets to protect \new{the network} from jamming attacks. The main challenges of developing defensive network flow are: (1) applying deceptive packets may increase the risk of congestion and incur extra delay for the delivery of real packets; and (2) the source node has a limited capacity to generate and transfer packets, requiring a balance between real and fake flows. To mitigate this adverse effect, the authors designed a two-stage game model to obtain deception strategies at pure-strategy Stackelberg equilibrium. They considered two types of source nodes: (1) \emph{selfish} nodes aiming to maximize its own utility; and (2) \emph{altruistic} nodes considering the congestion of other sources when choosing a flow rate. Their results proved that altruistic node behavior improves the overall utility of the sources. Similarly, \citet{zhu2012deceptive} considered a single source selecting routing paths for real and deceptive flows. Before sending a deceptive flow, the source node is allowed to choose the rate of deceptive and real flow as well as the path of the deceptive flow. This work introduced the solution concepts, such as the path Stackelberg equilibrium (PSE), the rate Stackelberg equilibrium (RSE), and their mixed strategy counterparts of the game. Their results proved that there exist such equilibria. \citet{anjum2020optimizing} designed a deceptive network flow system, called {\em Snaz}, to mislead the attacker performing reconnaissance attacks. They model the interaction between the attacker and the defender with a two players non-zero-sum Stackelberg game. \citet{sayin2019deception} proposed a deceptive signaling game framework to deal with APT attacks in cyber-physical systems.  Under the attacks aiming to obtain system intelligence via scanning or reconnaissance attacks, this work crafted bait information to lure the attackers. This work leveraged the concept of game-theoretic hierarchical equilibrium and solved a semi-definite programming problem where the defender does not have perfect information by considering partial or noisy observations or uncertainty towards the attacker's goal.

\vspace{1mm}
\noindent {\bf Pros and Cons}: Deceptive network flow is a new defensive deception approach to effectively mitigate some malicious actions which cannot be defended by traditional defensive deception technologies, such as malicious network flow scanning and fingerprint. However, it is highly challenging to balance fake and real network flows that minimizes any additional network performance degradation.

\begin{table*}[th!]
    \centering
    \caption{\new{Summary of Game-Theoretic-based Defensive Deception Techniques}}\label{tab:gt-summary}
\scriptsize
\vspace{-2mm}
    \begin{tabular}{|P{0.5cm}|P{1.6cm}|P{2cm}|P{2.7cm}|P{2.5cm}|P{1.5cm}|P{2.5cm}|P{1.3cm}|}
    \hline
         {\bf Ref.} & {\bf DD technique} & {\bf Goal} & {\bf Tactic} & {\bf Expected effect} & {\bf Main attacks} & {\bf Game type} & {\bf Domain } \\
\hline
\hline
\cite{Cho19-hgt}& Fake patch & Asset protection & Mimicking; False information; Lies & Luring; Confusing; Misleading & APT & Hypergame & No domain specified \\
\hline
\cite{ferguson2019game}& Deceptive signal & Asset protection; Attack detection & Mimicking; Decoying & Luring; Misleading & NC & Hypergame & No domain specified\\
\hline
\cite{kiekintveld2015game}& Honeypot & Asset protection; Attack detection&Mimicking; Decoying & Luring; Misleading& NC/Probing & General-sum & CPS \\
\hline
\cite{mao2019game}&Honeypot & Asset protection; Attack detection& Mimicking; Decoying & Luring; Misleading& APT & Bayesian & SDN \\
\hline

\cite{pawlick2015deception} & Honeypot & Asset protection; Attack detection& Mimicking; Decoying & Luring; Misleading& NC & Signaling & No domain specified\\
\hline
\cite{pawlick2015flip}& Deceptive signal & Asset protection&Mimicking; Masking; Lies; Decoying &Misleading; Luring; Confusing & APT & Three-player signaling & Cloud Web \\
\hline
\cite{clark2012deceptive} & Dummy packet generation &Asset protection &Mimicking; Masking &Misleading; Hiding; Confusing &  Jamming & Stackelberg & Wireless \\
\hline
\cite{zhu2012deceptive}& Deceptive network flow & Asset protection & Mimicking; Masking & Misleading; Hiding; Confusing & Jamming & Multi-stage stochastic & No domain specified \\
\hline
\cite{mohammadi2016Springer}& Fake avatar & Asset protection & Mimicking; Decoying & Asset protection & NC & Signaling & No domain specified\\
\hline
\cite{pawlick2018modeling} & Honeypot &Asset protection; Attack detection & Mimicking; Decoying & Luring; Misleading  & NC & Signaling & No domain specified \\
\hline
\cite{pibil2012game} & Honeypot & Asset protection; Attack detection & Mimicking; Decoying & Luring; Misleading  & NC & General-sum & No domain specified\\
\hline
\cite{la2016deceptive}& Honeypot & Asset protection; Attack detection & Mimicking; Decoying & Luring; Misleading  & NC & Bayesian & IoT\\
\hline
\cite{cceker2016deception}& Honeypot & Asset protection; Attack detection & Mimicking; Decoying & Luring; Misleading & DoS & Signaling & No domain specified\\
\hline
\cite{basak2019identifying}& Honeypot & Asset protection; Attack detection & Mimicking; Decoying & Luring; Misleading & APT & General-sum & No domain specified \\
\hline
\cite{xi2020hypergame}& Honeypot & Asset protection; Attack detection & Mimicking; Decoying & Luring; Misleading & APT & Static & IoT\\
\hline
\cite{durkota2015optimal} & Honeypot & Asset protection; Attack detection & Mimicking; Decoying & Luring; Misleading & NC & Stackelberg & No domain specified\\
\hline
\cite{wagener2009self}& Honeypot & Asset protection; Attack detection & Mimicking; Decoying & Luring; Misleading & NC & General-Sum & No domain specified\\
\hline
\cite{aggarwal2016cyber}& Honeypot& Asset protection; Attack detection & Mimicking; Decoying & Luring; Misleading & NC/probing & Non-cooperative sequential; incomplete information & No domain specified \\
\hline
\cite{aggarwal2017modeling}& Honeypot& Asset protection; Attack detection & Mimicking; Decoying & Luring; Misleading & NC/probing & Sequential &  No domain specified\\
\hline
\cite{anwar2020gamesec}& Deceptive signal  & Asset protection & Mimicking; Masking; Decoying & Misleading; Luring; Confusing & NC/Probing & Stochastic with POMDP & IoT \\
\hline
\cite{anwar2019game}&Honeypot & Asset protection; Attack detection & Mimicking; Decoying & Luring; Misleading & NC & Stochastic & IoT \\
\hline
\cite{anwar2020Honeypotbook}&Honeypot & Asset protection; Attack detection & Mimicking; Decoying & Luring; Misleading & NC & Stochastic & IoT \\
\hline
\cite{nan2019behavioral} & Honeypot & Asset protection; Attack detection & Mimicking; Decoying & Luring; Misleading & NC & Static  & IoT\\
\hline
\cite{el2018new}& Honeywebs &Asset protection &Decoying; Mimicking & Blending; Misleading; Luring& Web attack &Complete, imperfect game& Cloud Web\\
\hline
\cite{garg2007deception} & Honeynet & Asset protection; Attack detection & Decoying; Mimicking & Asset protection; Attack detection & NC & Zero-sum static & No domain specified \\
\hline
\cite{dimitriadis2007improving} & Honeypot & Asset protection; Attack detection & Decoying; Mimicking & Blending; Misleading; Luring & DDoS & Zero-sum static & Wireless \\
\hline

\cite{hespanha2000deception} & Obfuscation & Asset protection & Masking & Misleading; Confusing & NC & Non-cooperative stochastic & No domain specified \\
\hline
\cite{yin2013optimal}& Deceptive signal & Asset protection & Masking; Camouflaging; Mimicking & Hiding; Blending, Hiding; Misleading & Reconnaissance & Stackelberg & No domain specified \\
\hline
\cite{horak2017manipulating} & Deceptive signal & Asset protection; Attack detection & Misleading; Hiding; Mimicking; Decoying & Misleading; Hiding; Luring; Confusing; Blending & NC & Stochastic & No domain specified\\
\hline
\cite{bilinski2019you}& Deceptive signal & Asset protection; Attack detection & Misleading; Hiding; Mimicking; Decoying & Asset protection; Attack detection & NC & Stackelberg & No domain specified\\
\hline
\cite{chiang2018defensive} & Honey-X technologies &Asset protection; Attack detection &Misleading; Hiding; Mimicking; Decoying &Misleading; Hiding; Luring; Confusing; Blending & APT & General-sum  & SDN \\
\hline
\cite{huang2019dynamic}& Deceptive signal & Asset protection; Attack detection & Misleading; Hiding; Mimicking; Decoying & Misleading; Hiding; Luring; Confusing; Blending & APT & Dynamic Bayesian & No domain specified \\
\hline
\cite{casey2015compliance}& Honey surface & Asset protection & Mimicking; Decoying & Hiding; Misleading & Insider & Signaling & No domain specified \\
\hline
\cite{casey2016compliance}& Honey surface & Asset protection & Mimicking; Decoying & Hiding; Misleading & Insider & Signaling & No domain specified\\
\hline
\cite{thakoor2019cyber}& Fake objects & Asset protection & Mimicking; Decoying & Hiding; Misleading &  Recon. & General-sum  & No domain specified\\
\hline
\cite{Zhu14-sh}& Social honeypots & Asset protection; Attack detection & Decoying; Mimicking & Blending; Misleading; Luring & Social bots & Stackelberg & No domain specified \\
\hline
\cite{sayin2019deception}& Deceptive network flow & Asset protection&Mimicking; Masking &Misleading; Hiding; Confusing &  Recon./Probing & Non-zero-sum Stackelberg & CPS \\
\hline

\cite{al2019attacker}& Honeypot & Asset protection; Attack detection & Decoying; Mimicking & Blending; Misleading; Luring & APT  & POMDP & SDN \\
\hline
\end{tabular}
\end{table*}

\begin{table*}[th!]
    \centering
    \caption*{\new{Continued: Summary of Game-Theoretic-based Defensive Deception Techniques}
    }
\scriptsize
\vspace{-2mm}
    \begin{tabular}{|P{0.5cm}|P{1.6cm}|P{2cm}|P{2.7cm}|P{2.5cm}|P{1.5cm}|P{2.5cm}|P{1.3cm}|}
    \hline
         {\bf Ref.} & {\bf DD technique} & {\bf Goal} & {\bf Tactic} & {\bf Expected effect} & {\bf Main attacks} & {\bf Game type} & {\bf Domain } \\
\hline
\hline
\cite{al2020dynamic}&Honeypot & Asset protection; Attack detection & Decoying; Mimicking & Blending; Misleading; Luring & APT & POMDP & SDN\\
\hline
\cite{nan2020mitigation} & Deception signal & Asset protection; Attack detection & Decoying; Mimicking & Blending; Misleading; Luring & NC & Stackelberg & Wireless \\
\hline
\cite{wang2020intelligent}& Honeypot & Asset protection; Attack detection & Decoying; Mimicking & Blending; Misleading; Luring & APT & Q-learning-based & No domain specified\\
\hline
\cite{anwar2020honeypot}&Honeypot & Asset protection; Attack detection & Mimicking; Decoying & Luring; Misleading & NC & Zero-sum static & IoT \\
\hline
\cite{al2019online}& Fake nodes & Asset protection; Attack detection & Decoying& Hiding; Misleading &  APT & POMDP &No domain specified  \\
\hline

\cite{rahman2013game} & Fake objects & Asset protection & Mimicking; Decoying & Hiding; Misleading & Recon. & Signaling & No domain specified\\
\hline
\cite{learning2020gamesec}& Feature deception & Asset protection & Lies & Misleading & Reconnaissance & RL-based feature deception & No domain specified\\
\hline
\end{tabular}
\end{table*}

In Table~\ref{tab:gt-summary}, we summarized the five key aspects (i.e., deception techniques, categories, expected effects, attacks countermeasured, and application domain) of a given game-theoretic defensive deception techniques surveyed in this work.

\subsection{Empirical Game Experiments Using Human Subjects}
As we observed in Section~\ref{subsubsec:GTDD}, most deception game-theoretic works have been studied based on simulation testbeds. However, some cognitive decision making scientists also conducted empirical experiments using human subjects which are assigned as attackers or defenders to consider a deception game.  \citet{ cranford2020adaptive, cranford2019towards} conducted an empirical experiment to realize a signaling game between an attacker and a defender where the defender uses deceptive signals. \new{In this experiment,} the players are humans with limited cognition, reflecting bounded rationality in game theory. To measure the effectiveness of defensive deception techniques, \citet{ferguson2018tularosa} and \citet{ferguson2020empirical} also conducted an empirical experiment with 130 red team members as participants in a network penetration study at different deception scenarios. For instance, different types of decoy devices are described explicitly for the existence of deceptive defensive techniques to the participant. The results showed that when the participants are aware of the deception, they took much more time before taking any action. This shows that deception made them slower to move and attack. \citet{aggarwal2019hackit} developed a simulation tool, called {\em HackIt}, to study attacks in a network reconnaissance stage where participants studied the effect of introducing deception at different timing intervals. Their results showed that the attacker performed attacks on the honeypots more often than real machines. 

Pure game theorists may argue that this type of empirical game experiments is not a game-theoretic deception game. However, we discuss these empirical deception game studies to give a chance to be aware of human subjects-based empirical game experiments that can at least partly follow the structure of a traditional game theory, consisting of two or more players, their bounded rationality, their opponent strategies and corresponding utilities that can affect their decision making.

\section{Machine-Learning-Based Defensive Deception (MLDD) Techniques}
\label{sec:ml-dd}

ML-based applications become more popular than ever in various domains, including cybersecurity. ML techniques have been extensively adopted for automating attacks and learning system behaviors in the cyberdeception domain~\cite{al2019towards}. In this section, we first discuss the key steps to implement ML-based approaches and conduct an extensive survey on ML-based defensive deception techniques in the literature.

\subsection{Key Steps of Implementing ML-Based Defensive Deception} \label{subsec:steps-ml}

Like other ML-based applications, ML-based deception techniques also commonly used the following steps to detect malicious activities:
\begin{itemize}
\item \new{{\em Dataset Generation}: ML-based approaches have been mainly used in detecting attackers or identifying malicious activities based on the information collected in honeynets or honeypots~\cite{song2011statistical}.} 
\item {\em Dataset Collection}:  The success of ML techniques hugely depends on whether reliable datasets are available or not. However, obtaining reliable cybersecurity datasets is not trivial based on the following reasons: (i) such datasets are often confidential due to security reasons of a company or an organization owning the datasets; and (ii) there are inherent unknown attacks which make it difficult to obtain accurate and complete annotated datasets. Cybersecurity related datasets have been more available as more security applications are using ML-based approaches~\cite{snijders2012big}.
\item {\em Pre-Processing}: Datasets should be carefully pre-processed ahead of the learning step to eliminate unwanted noises from the dataset. This includes data cleaning, editing, and reduction~\cite{han2011data}. 

\item {\em Feature Extraction}: It is challenging to analyze complex data because of the large number of variables involved. The goal of feature extraction is to construct a combination of the relevant variables to describe the data with sufficient accuracy.
\item {\em Training}: Training dataset is a dataset of examples used during the learning process and is used to learn the parameters of the detector~\cite{han2011data,ron1998special}. 
\item {\em Testing}: A testing dataset is a set of examples used to evaluate the performance (i.e, detection accuracy) of a trained detector~\cite{ron1998special}. 
\item {\em Evaluation}: Independent and new data are used to evaluate the performance of a detector to avoid overfitting problems in the data~\cite{ron1998special}.
\end{itemize}

\begin{table*}[th!]
\scriptsize 
    \centering
    \caption{\new{Summary of Machine Learning-based Defensive Deception Techniques}}
    \label{tab:ml-summary}
    \vspace{-2mm}
\begin{tabular}{|P{0.5cm}|P{1.8cm}|P{2cm}|P{2.7cm}|P{2cm}|P{2cm}|P{2cm}|P{1.5cm}|}
    \hline
         {\bf Ref.} & {\bf DD technique} & {\bf Goal} & {\bf Tactic} & {\bf Expected effect} & {\bf Main attacks} & {\bf ML technique} & {\bf Domain} \\ \hline
\hline
\cite{nanda2016NFVSDN} & Honeypot & Attack detection & Mimicking & Attack detection & Anomaly network traffic & C4.5, Bayesian Network, Decision Table, and Na\"{i}ve-Bayes & No domain specified \\
\hline
\cite{badri2016uncovering}& Social honeypot & Attack detection & Mimicking & Luring &  Fake liker (crowdturfing) & Supervised ML & OSN \\
\hline
\cite{nisrine2016security}& Social honeypot & Attacker identification & Mimicking & Luring & Malicious profile & ML toolkit & No domain specified \\
\hline
\cite{krueger2012ACMSAI}& Honeypot & Asset protection; Attack detection & Mimicking; Decoying & Luring; Misleading & Call fraud; Identity theft & ML toolkit & No domain specified \\
\hline
\cite{lee2010uncovering} & Social honeypot &Identifying spammers  &Mimicking; Decoying & Luring & Social spammers & ML toolkit & OSN \\
\hline
\cite{lee2011seven} & Social honeypot &Identifying spammers & Mimicking; Decoying& Luring & Social spammers & ML toolkit & OSN \\
\hline
\cite{hofer2019model}& Decoy & Asset protection; Attack detection & Decoying; Mimicking & Blending; Misleading; Luring & APT & RNN & CPS \\
\hline
\cite{ben2012combining}& Bait-based Deception & Asset protection; Attack detection & Mimicking; Lies & Luring; Confusing & Insider & SVM &No domain specified \\
\hline
\cite{whitham2016minimising}& Bait-based Deception & Attack protection & Mimicking; Lies & Luring; Confusing & Zero-day & NLP & No domain specified\\
\hline
\cite{stringhini2010detecting}& Honey profile & Asset protection & Mimicking; Decoying & Hiding; Misleading & Spamming & SVM & OSN \\
\hline
\cite{whitham2017automating}& Bait-based Deception & Attack detection & Mimicking & Luring & Zero-day & NLP & No domain specified \\
\hline
\cite{abay2019using}& Decoy data & Attack detection & Lies  & Misleading & Data leakout & DL & No domain specified\\
\hline
\end{tabular}
\end{table*}

\subsection{\new{Common ML Techniques Used for Defensive Deception}}

\new{We discuss popular ML techniques used to develop various DD techniques in the literature. For the easy understanding of readers who may not have enough background on machine learning, we give the brief overview of the following ML techniques:}
\begin{itemize}
\item \new{{\em Support Vector Machine (SVM)}: SVMs can deal with a set of supervised learning models and be applied to solve classification and regression problems~\cite{cortes1995support}. SVM uses the hypothesis space of linear functions in a high dimensional feature space to analyze data. SVM is popular due to its simple training process. However, SVM is not efficient for large or highly noisy datasets}

\item \new{{\em $K$-Means}: This technique uses an iterative refinement to partition $n$ observations into $k$ clusters and guarantee that each observation belongs to a cluster with the nearest mean~\cite{wiki:kmeans}.} 

\item \new{{\em Expectation Maximization (EM)}: This is a statistical method to search for maximum likelihood parameters. It consists of two major steps: (1) an expectation step, which creates a function for the expectation of the likelihood evaluated using the current estimate for the parameters; and (2) a maximization step, which computes a new estimate of the parameters~\cite{moon1996expectation}}

\item \new{{\em Hierarchical Grouping}: This is a well-established classification method for cluster analysis which seeks to build a hierarchy of clusters~\cite{ward1963hierarchical}. For example, plants and animals may be grouped into a smaller number of mutually exclusive classes with respect to their genetic characteristics.}

\item \new{{\em Bayesian Network (BayesNet)}: This is a classifier that learns the conditional probability of each attribute from training data while assuming strong independence~\cite{friedman1997bayesian}. That is, the underlying probability model would be an independent feature model such that the presence of a particular feature of a class is unrelated to the presence of any other features. For example, BayesNets could represent the probabilistic relationships between diseases and symptoms.}

\item \new{{\em Decision Tree (DT)}: DT is a widely used approach for multistage and sequential decision making. The key idea of DT lies in breaking up a complex decision into a union of several simpler decisions, leading to the intended desired solution~\cite{safavian1991survey}.}

\item \new{{\em C4.5 Algorithm}: This is a method to generate a decision tree based on the concept of information entropy in which the DT generated by C4.5 is widely used for data classification~\cite{quinlan2014c4}.}

\item \new{{\em Na\"{i}ve-Bayes Algorithm}: This is a widely applied classification algorithm based on Bayesian Theory and  statistical analysis~\cite{domingos1997optimality}. This algorithm uses the prior probability and posterior probability of the training data to calculate the probability of type (posterior probability) of a given sample.}

\item \new{{\em Deep Neural Networks}: DNN is a neural network with at least one hidden layer and it has been used as a framework for deep learning. For each node (neuron), its output is a non-linear function of a weighted sum of its inputs. The DNN model is trained by back propagation, which changes the weight of each node. DNNs have been used for natural language processing, image recognition, and disease diagnosis.}
\end{itemize}

\subsection{ML-Based Defensive Deception} \label{subsec:MLDD}
\new{In this section, we discuss ML-based defensive deception techniques that have been discussed in the literature. We discuss the following DD techniques along with what ML techniques are used to develop them: Social honeypots, honey files, honeypots or decoy systems, bait-based deception, fake services, obfuscation, and decoy data. }
\subsubsection{Social Honeypots} 
In online social networks (OSNs), the so-called {\em social honeypots}, as good bots in contrast to bad bots, have been studied by creating social network avatars.

\citet{lee2010uncovering} proposed an ML-based mechanism to discover social spammers. The authors used social honeypots as a defensive deception technique to monitor spammer activities. The data from spam and legitimate profiles are used to improve an ML-based classifier. Thus, the system can classify spam profiles automatically. The paper compared true positive rate (TPR) and false negative rate (FNR) of various ML-based classifiers to show the effectiveness of the proposed mechanism. \citet{lee2011seven} developed social honeypots to detect content polluters in Twitter. Sixty social honeypot accounts are created to follow other social honeypot accounts and post four types of tweets to each other. This work categorized the collected user features into nine classes based on the Expectation-Maximization (EM) algorithm. They classified the content polluters based on Random Forest and enhanced the results by using standard boosting and bagging and different feature group combinations. 

\citet{nisrine2016security} developed social honeypots to discover malicious profiles. The authors employed feature-based strategies as well as honeypot feature-based strategies to collect data of the malicious profiles and analyzed the collected data based on a suite of ML algorithms provided by Weka ML toolkit (e.g., logistic regression, Bayes  classifier, support vector machine (SVM), $K$-Means, Expectation Maximization, or hierarchical grouping). \citet{zhu2015fighting} introduced the concept of `active honeypots' as active Twitter accounts that can capture more than 10 new spammers every day. They identified 1,814 accounts from the Twitter and examined the key features of active honeypots. In addition, using a suite of ML algorithms, such as SVM, logistic regression, J48 Tree, Bagging, and AdaBoost M1, the authors examined the impact of unbalanced datasets on the detection accuracy of the different ML algorithms. 

\citet{badri2016uncovering} also used social honeypot pages to collect fake likers on Facebook where fake likes are provided by paid workers. The authors obtained four types of user profiles and behavioral features as the distinctive patterns of fake likers. This work evaluated the robustness of their ML-based detection algorithms based on synthetic datasets which have been modified to reflect individual and coordinated attack models. \citet{yang2014taste} proposed a passive social honeypot to collect a spammer's preferences by considering various behaviors of the social honeypot. The considered tweet design features include tweet frequency, tweet keywords, and tweet topics as well as the behaviors of famous users' accounts and application installation. The authors conducted in-depth analysis on what types of social honeypot features can introduce a higher rate of collecting social adversaries and enhanced the social honeypots based on the result.  The improved honeypot has shown 26 times faster than a normal social honeypot in collecting social adversaries based on the evaluation using a random forest classifier.

\vspace{1mm}
\noindent {\bf Pros and Cons}: The works discussed in~\cite{nisrine2016security,lee2010uncovering,lee2011seven,yang2014taste} mainly relied on luring deception techniques to attract spammers in OSNs and collect more attack intelligence particularly in terms of novel behavioral characteristics of spammers.  Some limitations of the existing social honeypots include: (i) the current efforts are mainly to detect spammers, not other social network attacks (e.g., cybergroomers, human trafficking, cyberstalking, or cyberbullying); (ii) most features are based on user behaviors, but not really network topological features as an attacker's characteristics; (iii) most detectors are applied on static datasets to evaluate the detection of spammer. This implies that there is no theoretical simulation framework that considers dynamic interactions between an attacker and a defender; and (iv) there is a lack of studies investigating the effectiveness of social honeypots in terms of how quickly more attack intelligence can be collected and what types of attack intelligence are collected.

\subsubsection{Honey Profiles} \citet{stringhini2010detecting} analyzed 900 honey profiles for detecting spammers in three online social networks, including MySpace, Facebook, and Twitter. The authors collected users' activity data for a year where the user datasets include both spammers' and legitimate users' profiles and the honey profiles are spread in different geographic networks. Further, this work captured both spam profiles and spam campaigns based on the shared URL using machine learning techniques (e.g., SVM).

\vspace{1mm}
\noindent {\bf Pros and Cons}: Honey profiles allow identifying a large set of attackers, such as a large-scale spam bots, which lead to the detection of large-scale and coordinated campaigns. However, the existing spam detectors using honey profiles is not generic and needs to be tailored depending on a different social media platform.

\subsubsection{Honeypots or Decoy Systems}  \citet{nanda2016NFVSDN} used a suite of machine learning algorithms to train historical network attack data in software-defined networks for the development of quality honeypots. This ML-based method is to identify potential malicious connections and attack destinations. The authors employed four well-known ML algorithms, including C4.5, Bayesian Network (BayesNet), Decision Table (DT), and Na\"{i}ve-Bayes to predict potential victim hosts based on the historical data.  To prevent call fraud and identity theft attacks, \citet{krueger2012ACMSAI} proposed a method called {\em PRISMA} (PRotocol Inspection and State Machine Analysis).  This method is developed to infer a functional state machine and a message format of a protocol from only network traffic.  The authors experimented to validate the performance of the PRISMA based on three real-world network traces datasets, including 10,000 to 2 million messages in both binary and textual protocols. Their experiments proved that the PRISMA provides the capability to simulate complete and correct sessions with the learned models and to execute different states for malware analysis.

\citet{hofer2019model} discussed the attributes of cyber-physical decoys to design a system with these attributes to develop deception decoys.  The authors integrated the deception decoys into real systems to make them harder to detect and more appealing as targets. To increase the fidelity of the added decoy devices to the physical system, three attributes, a protocol, variables, and a logic of the deployed decoy devices, are maintained.  The authors trained a recurrent neural network (RNN) using a dataset collected for a year to learn such system attributes.

\new{Some defensive deception games have been studied by considering game theoretic reinforcement learning (RL) where RL is considered in formulating players (i.e., attackers or defenders)'s utilities.  That is, in the RL-based game formulation, players use an RL to identify an optimal strategy where the RL's reward function considers gain and loss based on the player's belief towards an opponent's move.  RL-based deception games are studied in \cite{wang2020intelligent,al2019online}. \citet{al2019online} proposed an online deception approach that designs and places network decoys considering scenarios where a defender's action influences an attacker to dynamically change its strategies and tactics while maintaining the trade-off between availability and security. The defender maintains a belief consisting of a security state while the resultant actions are modeled as Partially Observable Markov Decision Process (POMDP).  The defender uses an online deception algorithm to select actions based on the observed attacker behavior using a POMDP model. This embedded reinforcement learning (RL)-based model assumes that the defender belief about the attacker's progress is observed through an network-based intrusion detection system (NIDS). The defender hence takes actions that attract the attacker toward decoy nodes.  \citet{wang2020intelligent} identified an optimal deployment strategy for deception resource, such as honeypots.  Specifically, the authors developed a Q-learning algorithm for an intelligent deployment policy for deception resources to dynamically place deception resources as the network security state changes.  This work considered an attacker-defender game by analyzing an attacker's strategy under uncertainty and a defender's strategies with several deployment location policies.  This work used RL with a Q-learning training algorithm that identifies an optimal deployment policy for deception resources.}

\vspace{1mm}
\noindent {\bf Pros and Cons}: In honeypots, ML-based approach is mainly used to detect attacks based on the attack intelligence gathered in honeypots. Hence, this research direction is like developing an intrusion detection mechanism, rather than creating better quality defensive deception techniques. Various types of ML-based techniques can be highly leveraged in order to create real-like honeypots and its effectiveness can be measured based on the volume of attackers caught in honeypots and the diversity of attacker types.   

\subsubsection{Bait-based Deception} \citet{ben2012combining} developed a bait-based deception to improve the detection of impersonation attacks by using the features combining a user behavior profiling technique with a baiting deception technique. This work used highly crafted and well-placed decoy documents to bait attackers and deployed a detector that uses SVM to model the user's search behavior. \new{\citet{whitham2016minimising} used bait-based deception technique to study and evaluate honey files in military networks. The main challenge in military networks is the need to utilize third party resources such as cloud services. Third party servers are directly connected to the Internet and may store sensitive material and information which represents a security challenge. The author proposed three new automated honey file designs that minimize the replication of classified material, yet still remain enticing to malicious software or user driven searches. Moreover, this work presented an NLP-based content generation design for honey files.}

\vspace{1mm}
\noindent {\bf Pros and Cons}: Although bait-based deception can contribute to increasing intrusion detection by collecting more intelligence during the time of the deception, there is no performance guarantees based on the bait-based deception because the attackers may not be interested in them. In addition, to more effectively attract attackers, if more important information is used as a bait, it may introduce some vulnerability or risk when intelligent attackers can derive clues of system vulnerabilities based on the bait. Thus, mixing real with fake information to avoid too high a risk can provide an alternative, such as semi-bait-based deception.

\subsubsection{Fake Services} To increase the effectiveness and efficiency of defensive deception techniques, real systems have been used to deceive attackers. 
ML-based fake system modification is used as a defensive deception mechanism~\cite{al2019attacker,al2020dynamic} using POMDP to consider uncertainty in learning attack behavior. The authors designed a deception server that can modify IP addresses of fake network via DHCP server, DNS server, and forwarding ARP requests by appropriate flow rules to the deception server. In addition, the deception server is designed to modify the requests and sent it back to the requesting host. \citet{learning2020gamesec} considered the changes of system features and configurations to deceive an attacker after learning the attacker's preferences and its behavior from attack data.

\vspace{1mm}
\noindent {\bf Pros and Cons}: Proving fake services to attackers can effectively increase attack cost or complexity which can delay launching attacks or make the escalation of attack fail.  However, this technique can be deployed based on the assumption of accurate detection of intrusions. As the current intrusion detection mechanisms suffer from high false positives in practice, if a normal user is treated as an intrusion and fake services are provided to the user, it hinders normal availability of service provisions to legitimate users.




\subsubsection{Decoy Data} \label{subsubsec:decoy-data} Machine and deep learning (DL) algorithms have played a key role in developing decoy data, creating decoy objects or honey information. \citet{abay2019using} took a decoy data generation approach to fool attackers without degrading system performance by leveraging DL. To be specific, their approach employs unsupervised DL to form a generative neural network that samples HoneyData. 

\vspace{1mm}
\noindent {\bf Pros and Cons}: Since the defender launches false information injection (or data disruption) attack while an attacker stays in a system, this technique will 'scare off' attackers performing highly detectable attacks. However, it can also be disruptive to ordinary users if fake or decoy information is allowed to remain in the system.

\new{As we observed in the extensive discussions of ML-based defensive deception techniques above, ML-based defensive deception has been addressed mainly with the following aims: (1) Protecting system components by hiding real components or creating fake things (i.e., honey-X) that look like real; and (2) Detecting attacks (or identifying attackers) by creating fake objects (e.g., honeypots) and accordingly attract the attackers to collect attack intelligence.  Therefore, what and how ML techniques are used is closely related to the quality of defensive deception techniques whose success is closely related to how well it can deceive attackers, representing the quality of deception.}

In Table~\ref{tab:ml-summary}, we summarized the five key aspects (i.e., deception techniques, categories, expected effects, attacks countermeasured, and application domain) of a given ML-based defensive deception techniques surveyed in this work.

\section{Attacks Countermeasured by Defensive Deception Techniques} \label{sec:attacks}
A variety of attacks have been countered by existing defensive deception techniques as follows: 
\begin{itemize}[leftmargin=*]
\item {\em Scanning (or reconnaissance) attacks}~\cite{mao2019game,xu2016modeling,bilinski2019you,chiang2018defensive,avery2018CS,anjum2020optimizing,sayin2019deception,al2019attacker}: An outside attacker may aim to obtain a target system's information and identify vulnerable system component to penetrate into the target system. This attack is often considered as part of the cyber kill chain in APT attacks. In the literature, the scanning attacks are categorized as two types as follows:
\begin{itemize}
\item {\em Passive monitoring attacks}~\cite{xu2016modeling,bilinski2019you,anjum2020optimizing}: Leveraging compromised nodes (e.g., routers or switches), attackers can scan traffics passing through the nodes and analyze packets to gather information about hosts. Passive monitoring can make the attackers achieve active nodes' discovery, OS, roles, up-time, services, supporting protocols, and IP network configuration~\cite{montigny2004passive}. However, passive monitoring cannot identify services that do not run on well-known ports or are misdirected without protocol-specific decoders~\cite{bartlett2007understanding}.
\item {\em Active probing attacks}~\cite{mao2019game,chiang2018defensive,avery2018CS,sayin2019deception,al2019attacker}: Through sending probing packets to a potential target, the attacker can obtain the machine's information, includes OS, opened ports, and installed application version.  Although active probing can allow the attackers to collect more system configurations, it is relatively easy to be detected by traditional defensive technologies, such as IDS~\cite{bartlett2007understanding}. 
\end{itemize}

\item {\em Network fingerprinting}~\cite{rahman2013game}: To analyze a target network and search for an ideal target node, an attacker can use fingerprint tools to identify specific features of a network or a device, such as a network protocol or an OS running on a remote machine. Common fingerprint tools include Nmap~\cite{nmap}, P0f~\cite{Zalewski14}, and Xprobe~\cite{arkin2002xprobe}. Some deception techniques, such as deceptive network flows and honeypots, focuses on addressing this kind of attacks.
    
\item {\em (Distributed) Denial-of-Service} (DoS)~\cite{cceker2016deception,dimitriadis2007improving}: 
DoS attacks are mainly to make legitimate users be denied from proper services. This attack can overwhelm or flood network flow to a targeted machine to disrupt normal functions or communications.  A common DoS attack is ping flooding by leveraging ICMP (ping) packets to overwhelm a target~\cite{peng2007survey}. In distributed environments, distributed DoS (DDoS) flooding attacks can be performed with the intent to block legitimate users from accessing network resource(s). With flooding packets to a target, DDoS attackers can exhaust network bandwidth or other resources, such as a server's CPU, memory or I/O bandwidth.
Unlike DoS attacks, DDoS attackers can remotely control a large number of devices as its botnet can continuously send a large amount of traffics to a target to block the normal functions and services~\cite{zargar2013survey}.

\item {\em Malware}~\cite{krueger2012ACMSAI}: Malware refers to any software aiming to introduce some damage to a system component, such as server, client, or network. Honeypot-based deception approaches have been popularly used for malware analysis~\cite{krueger2012ACMSAI}.

\item {\em Privacy attacks}~\cite{Shokri15}: An attacker can identify private information while training and processing data to develop defensive deception techniques. Obfuscation is often used to defend against privacy attacks. 

\item {\em Advanced persistent threats} (APT)~\cite{Cho19-hgt, pawlick2015flip, pawlick2018modeling, la2016deceptive, huang2019dynamic,learning2020gamesec, hofer2019model}: An APT attack is the most well-known sophisticated attack performing different types of attacks in the stages of the cyber kill chain (CKC)~\cite{okhravi2013survey}. In particular, game-theoretic defensive deception research has considered the APT attack but mostly focused on the attacks during the reconnaissance stage.

\item {\em Spamming}~\cite{lee2010uncovering, lee2011seven, stringhini2010detecting, zhu2015fighting, yang2014taste}: Online users may receive unsolicited messages (spam), ranging from advertising to phishing messages~\cite{Rathore17-is-survey}.

\item {\em Malicious or fake profiles}~\cite{nisrine2016security} (a.k.a. Sybil attacks): OSN attackers can create a large number of fake identities to achieve their own selfish goal based on others' personal information, such as e-mail, physical addresses, date of birth, employment date, or photos.

\item {\em Fake likers by crowdturfing}~\cite{badri2016uncovering}: Human attackers, paid by malicious employers, can disseminate false information to achieve the malicious employers' purposes via crowdsourcing systems~\cite{wang2012serf}. This is called {\em crowdturfing} and mostly aimed to spread fake information to mislead people's beliefs. 

\item {\em Loss of confidentiality, integrity, and availability (CIA)} ~\cite{casey2016compliance, cranford2020adaptive, cranford2019towards}: An insider attacker, called a {\em traitor}~\cite{salem2008survey}, can leak out confidential information to the outside as a legitimate user. Further, by using its legitimate status with the authorization of a target device, it can perform attacks (e.g., illegal access, modification or transformation of confidential information) that can lead to loss of integrity and availability.

\item {\em Impersonation attacks}~\cite{ben2012combining}: An inside attacker can masquerade a legitimate user's identity to access resources in a target system component~\cite{salem2008survey}.

\item {\em Jamming attack}~\cite{clark2012deceptive,zhu2012deceptive,nan2020mitigation}: A jamming attack can be seen as a subset of DoS attacks. The difference is that jamming attacks mainly focus on incurring traffics in wireless networks while the DoS attacks can be applicable in all networks. The jamming attack is relatively simple to archive. By keeping flooding packets, a jamming attacker, also named {\em jammer}, can effectively block the communication on a wireless channel, disrupt the normal operation, cause performance issues, and even damage the control system~\cite{grover2014jamming}. 


\item {\em Node compromise (NC)}~\cite{ferguson2019game,kiekintveld2015game,pawlick2015deception,pawlick2018modeling,pibil2012game,la2016deceptive,basak2019identifying,xi2020hypergame,durkota2015optimal,wagener2009self,aggarwal2016cyber,aggarwal2017modeling,anwar2020gamesec,anwar2019game,anwar2020Honeypotbook,nan2019behavioral,nan2020development,garg2007deception,horak2017manipulating,bilinski2019you,anwar2020honeypot}: Some research does not specify the details of attack process. The authors only use ``device compromising'' to represent an attack. Some research discusses that an attacker can probe a target before attacking~\cite{ferguson2019game,kiekintveld2015game,pibil2012game,la2016deceptive,xi2020hypergame,wagener2009self,aggarwal2016cyber,aggarwal2017modeling,anwar2020gamesec,anwar2019game,anwar2020Honeypotbook,nan2019behavioral,nan2020development,garg2007deception,anwar2020honeypot} while others only discuss the attacking actions~\cite{pawlick2015deception,pawlick2018modeling,basak2019identifying,durkota2015optimal,horak2017manipulating,bilinski2019you}.
\item {\em Web attack}~\cite{el2018new}: An attacker can leverage existing vulnerabilities in web applications to gather sensitive data and gain unauthorized access to the web servers. 
\item \new{{\em Zero-day attack}~\cite{whitham2016minimising, whitham2017automating}: An attacker can exploit the vulnerability period where an unknown vulnerability is not mitigated by a defense system.  For example, before a vulnerability is patched by the system, the attacker can exploit the vulnerability to launch the attack, which is called zero-day attack. Often times, defensive deception along with moving target defense (e.g., network address or topology shuffling~\cite{djamaluddin2018web}) can mitigate this attack effectively.}
\end{itemize}

\begin{figure}[htb]
\vspace{-3mm}
\centering
\begin{tikzpicture}
[pie chart,
slice type={Node Compromise}{blue!50},
slice type={Scanning or Reconnaissance}{blue},
slice type={APT}{orange},
slice type={Loss of CIA}{brown},
slice type={Jamming attack}{cyan},
slice type={DoS or DDoS}{black},
slice type={Network Fingerprinting}{brown!30},
slice type={Privacy attack}{black!30},
slice type={Web attack}{yellow},
pie values/.style={font={\small}},
scale=2]
\pye[xshift=0.6cm,values of coltello/.style={pos=1.1}]{1994}{
21/Node Compromise,
7/Scanning or Reconnaissance,
5/APT,
3/Loss of CIA,
3/Jamming attack,
2/DoS or DDoS,
1/Network Fingerprinting,
1/Privacy attack,
1/Web attack}
\legend[shift={(2cm, 1.2cm)}]{{Node Compromise (NC)}/Node Compromise, {Scanning or Reconnaissance}/Scanning or Reconnaissance,{APT}/APT,{Loss of CIA}/Loss of CIA, {Jamming attack}/Jamming attack, {DoS or DDoS}/DoS or DDoS, {Network Fingerprinting}/Network Fingerprinting, {Privacy attack}/Privacy attack, {Web attack}/Web attack}
\end{tikzpicture}
\caption{Types and frequency of attacks considered in the existing game-theoretic defensive deception approaches.}
\label{fig:attack_frequency-gt}
\end{figure}
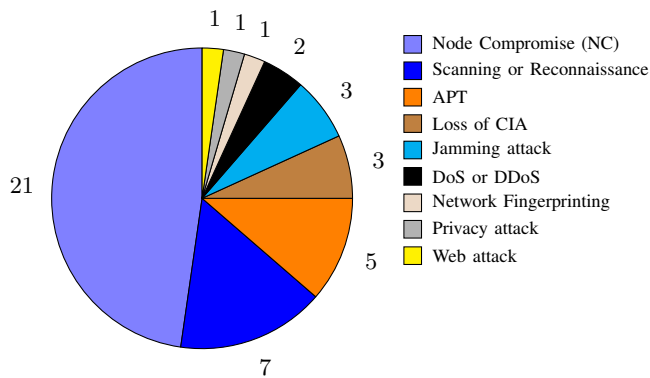

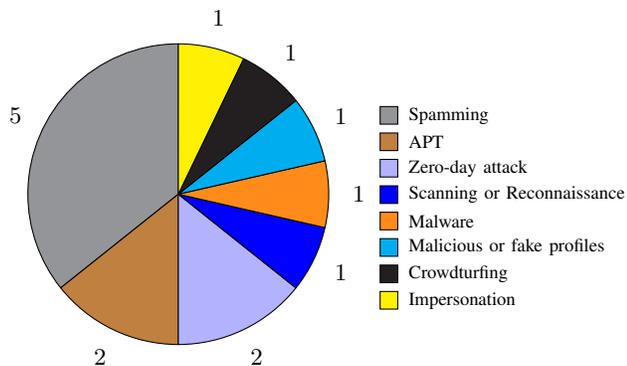
\begin{figure}[htb]
\vspace{-3mm}
\centering
\begin{tikzpicture}
[pie chart,
slice type={Spamming}{Black!50},
slice type={APT}{brown},
slice type={Zero-day attack}{blue!30},
slice type={Scanning or Reconnaissance}{blue},
slice type={Malware}{orange!90},
slice type={Malicious or fake profiles}{cyan},
slice type={Crowdturfing}{Black},
slice type={Impersonation}{yellow},
pie values/.style={font={\small}},
scale=2]
\pye[xshift=0.6cm,values of coltello/.style={pos=1.1}]{1994}{
5/Spamming,
2/APT,
2/Zero-day attack,
1/Scanning or Reconnaissance,
1/Malware,
1/Malicious or fake profiles,
1/Crowdturfing,
1/Impersonation}
\legend[shift={(2cm, 0.7cm)}]{{Spamming}/Spamming,{APT}/APT, {Zero-day attack}/Zero-day attack, {Scanning or Reconnaissance}/Scanning or Reconnaissance,{Malware}/Malware,{Malicious or fake profiles}/Malicious or fake profiles,{Crowdturfing}/Crowdturfing,{Impersonation}/Impersonation}
\end{tikzpicture}
\caption{Types and frequency of attacks considered in the existing machine-learning defensive deception approaches.}
\label{fig:attack_frequency-ml}
\vspace{-3mm}
\end{figure}

In Figs.~\ref{fig:attack_frequency-gt} and \ref{fig:attack_frequency-ml}, we summarized the number of game-theoretic or ML-based defensive deception (DD) techniques that handled each type of attacks discussed in this paper. \new{Note that the surveyed papers in this work were collected based on the following keywords: cyber deception, defensive deception, game theoretic defensive deception, and machine learning-based defensive deception. The surveyed papers in this work were published from 1982 to 2020.} As shown in Fig.~\ref{fig:attack_frequency-gt}, the majority of game-theoretic DD techniques considered node capture or compromise most and scanning (or reconnaissance) attack as the second most. As illustrated in Fig.~\ref{fig:attack_frequency-ml}, ML-based DD techniques considered spamming or malicious/fake profile attacks most as most ML-based approaches are mainly used to detect attacks in honeypots. 

\section{Application Domains for Defensive Deception Techniques} \label{sec:app-domains}

In this section, we provide how existing defensive deception techniques have been studied in different network environments or platforms. To be specific, we described the characteristics of each network environment and corresponding defensive deception techniques using game-theoretic or ML-based approaches, attacks considered, and metrics and experiment testbeds used.

\subsection{\new{Non-Domain Specific Applications}}

An enterprise network is a common system that is homogeneously configured to be operated in a static configuration~\cite{Kandula09}. Hence, an attacker can easily plan and perform its attack successfully and accordingly penetrate into the system without experiencing high difficulty but using dynamic strategies to defeat defense strategies.

\subsubsection{DD Techniques} The common defensive deception (DD) techniques used in this network environment include various types of honey information, such as fake honey files~\cite{whitham2016minimising} or honeypots~\cite{ pawlick2015deception, pawlick2018modeling, pibil2012game,cceker2016deception, Zhu14-sh, Rowe06-fake-honeypots}.

\subsubsection{Main Attacks} In enterprise networks, the following attacks have been countered by game-theoretic (GT) or ML-based DD, including insider threat and sophisticated adversaries, such as APTs~\cite{horak2017manipulating}, zero-day attacks~\cite{whitham2016minimising, whitham2017automating}, reverse engineering attacker using security patch~\citet{Avery17}, or
DoS attack~\cite{cceker2016deception}.

\subsubsection{Key GT and ML Methods} For game-theoretic DD techniques deployed in this network environment, signaling games~\cite{pawlick2015deception,cceker2016deception,  pawlick2018modeling} or Stackelberg game~\cite{Zhu14-sh} have been commonly used. Some different game types, such as Metagames/Expected Utility~\cite{pibil2012game} or
Bayesian Game using Bayesian equilibrium~\cite{cceker2016deception} have been also considered to develop DD techniques in this network context. For ML-based DD techniques, Natural Language Processing (NLP) techniques~\cite{whitham2016minimising, whitham2017automating} have been used to develop honey files. 

\subsubsection{Pros and Cons} Majority of game-theoretic DD approaches has considered an enterprise network without providing the details of a network model with the aim of examining a proposed game-theoretic framework with solid theoretical analysis, such as identifying optimal strategies or solutions based on Nash Equilibria. However, highly theoretical game-theoretic DD approaches may not provide concrete details on how to design DD techniques considering the characteristics of a given network environment in terms of resource availability, system security and performance/services requirements, or network and environmental dynamics. In addition, some general game-theoretic DD approaches do not specify particular DD techniques, but can provide a general idea of modeling a DD technique. However, this lacks details, showing limited applicability in real systems. Further, enterprise networks are mostly highly complex systems which may need to deal with a wide range of attack types. Due to the abstract nature of game-theoretic approaches proposed for the enterprise networks, they \new{do not} provide how to model and simulate specific types of attacks.

\subsection{\new{General} Cyber-Physical Systems (CPSs)} \label{subsec:cps}

A CPS is a system that can provide communications between humans via cyber capabilities and physical infrastructure or platforms~\cite{Poovendran10}. The CPS has been advanced with the cyber capabilities of communications, networking, sensing, and computing as well as physical capabilities with materials, sensors, actuators, and hardware. The CPS is uniquely distinguished from other platforms due to the presence of both cyber and physical aspects of systems and their coordination~\cite{Poovendran10}. \new{CPSs include wireless sensor networks, Internet of Things (IoT), software-defined networks (SDNs), and industrial control systems (ICSs)~\cite{kathiravelu2017sd,Serpanos18-cps}. Although we discuss each of these networks in separate sections, we include this section particularly to discuss defensive deception techniques that are designed for `general CPSs' without specifying a particular platform.}

\subsubsection{DD Techniques} Although defensive deception techniques have been deployed in CPS environments, their applicability seems not limited to only CPS environments, rather applicable in any environments. We found bait information and honeypots are applied to the CPS environments for introducing confusion or uncertainty to attackers or luring them to honeypots for collecting attack intelligence~\cite{ kiekintveld2015game, sayin2019deception,hofer2019model}.

\subsubsection{Main Attacks} game-theoretic or ML-based DD techniques in this environment mainly handled node compromise~\cite{kiekintveld2015game}, scanning or reconnaissance attacks~\cite{sayin2019deception}, or APT attacks~\cite{hofer2019model}.

\subsubsection{Key GT and ML Methods} A signaling game theory was used for a defender to identify an optimal strategy using DD techniques~\cite{sayin2019deception}. In addition, how to select honeypots to maximize confusion or uncertainty to attackers is also studied in~\cite{kiekintveld2015game}. To create decoy devices that look like real, RNN (Recurrent Neural Networks) was also used based on observations of device behaviors for a year in a CPS~\cite{hofer2019model}.

\subsubsection{Pros and Cons} Since a CPS is a popular network environment and commonly used in real systems, developing DD techniques for the CPS can have high applicability in diverse CPS contexts. However, the DD techniques developed for the CPS \new{did not} really consider much about its unique characteristics, such as the features of having both cyber and physical aspects, and challenges of the environment itself. Hence, we \new{do not} really observe much differences between the CPS's DD and the enterprise network's DD techniques. In addition, physical honeypots are sometimes recommended when dealing with highly intelligent attackers. However, such honeypot deployment and maintenance often come with higher deployment and management cost.

\subsection{Cloud-based Web Environments} \label{subsec:cloud}

Cloud-based web applications become more common and popular than ever to deal with critical tasks which bring security as a prime concern. Defensive deception is one of promising directions to defend against web attacks~\cite{Efendi19-cloud}. Defensive deception techniques can be utilized to defend against advanced web attacks that cannot be well handled by existing signature or anomaly-based intrusion detection or prevention mechanisms~\cite{Efendi19-cloud}. 

\subsubsection{DD Techniques} In  this environment, honey-X~\cite{el2018new} or fake signals~\cite{pawlick2015flip} are used as defensive deception techniques, such as honeywebs using honey files, honey tokens, decoy resources, or honeypots. 

\subsubsection{Main Attacks} Game-theoretic or ML-based DD techniques in cloud environments have mainly considered general web attacks~\cite{el2018new}, or multi-staged APT attacks~\cite{pawlick2015flip}.

\subsubsection{Key GT and ML Methods} Signaling games with complete, imperfect information have been used to model attack-defense games~\cite{pawlick2015flip,el2018new}. 

\subsubsection{Pros and Cons} While game-theoretic approaches proposed a generic game-theoretic DD framework~\cite{pawlick2015flip,el2018new}, ML-based obfuscation technique is fairly unique as it aims to defend against privacy attacks to deal with adversarial ML examples. However, we \new{do not} really observe concrete design features of a given DD technique only for the cloud computing environment. 

\subsection{Internet of Things (IoT)}  \label{subsec:iot}

IoT network environments become highly popular and have been recognized as one of CPSs with the capability to provide effective services to users. IoT has been also specialized for particular domains such as Internet-of-Battle-Things (IoBT)~\cite{Kott16}, Industrial-Internet-of-Things (IIoT)~\cite{Pinto17-IIoT}, or Internet-of-Health-Things (IoHT)~\cite{Rodrigues18-IoHT}.  In addition, IoT embraces a large number of CPS network environments, such as wireless sensor networks (WSNs), mobile ad hoc networks (MANETs), or smart city environments consisting of sensors and other intelligence machines~\cite{kocakulak2017overview}. IoT has been popular considered as a main platform of deploying cyberdeception technologies~\cite{alshammari2020deception}. 

\subsubsection{DD Techniques} Similar to other environments, honeypot enabled IoT networks have been mainly considered by solving the optimal deployment of honeypots~\cite{la2016deceptive,cceker2016deception}.

\subsubsection{Main Attacks}  In IoT environments, game-theoretic or ML-based DD techniques aimed to defend against reconnaissance and probing attacks, DoS attacks~\cite{cceker2016deception}, packet dropping attacks~\cite{cceker2016deception}, or APT attacks~\cite{la2016deceptive,xi2020hypergame}.

\subsubsection{Key GT and ML Methods} Bayesian games with incomplete information and meta games are considered for a player who is unsure of a type of attackers~\cite{ la2016deceptive}. IoTs in battlefields is referred to IoBTs where deception games introduced in \cite{xi2020hypergame,anwar2020gamesec,anwar2019game,anwar2020Honeypotbook, nan2019behavioral,anwar2020honeypot}


\subsubsection{Pros and Cons} Since honeypots are fake nodes mimicking the behavior of a regular node, adding a honeypot \new{does not} change the hierarchy of the IoT network or the interface of IoT gateways. A game-theoretic honeypot technique is the only technique applied to this domain. However, the IoT naturally can generate a large amount of data for ML-based DD techniques which can create real-like decoys or enhance detecting inside and outside attackers.

\begin{table*}[th!]
    \centering
    \caption{Application Domains of game-theoretic or Machine Learning-based Defensive Deception Techniques}
    \label{tab:app-domains}
    \vspace{-2mm}
    \begin{tabular}{|P{2.5cm}|P{2.5cm}|P{2.5cm}|P{2.5cm}|P{2.5cm}|P{3cm}|}
    \hline
         {\bf Application domain} & {\bf Defensive deception techniques} & {\bf Main attacks} & {\bf Key GT/ML techniques} & {\bf Pros} & {\bf Cons} \\
    \hline
    \new{Non-domain specific environment}:
    GT-Based~\cite{Cho19-hgt,ferguson2019game,pawlick2015deception,zhu2012deceptive,mohammadi2016Springer,pawlick2018modeling,pibil2012game,cceker2016deception,basak2019identifying,durkota2015optimal,wagener2009self,aggarwal2016cyber,aggarwal2017modeling,garg2007deception,hespanha2000deception,yin2013optimal,horak2017manipulating,bilinski2019you,huang2019dynamic,casey2015compliance,casey2016compliance,thakoor2019cyber,Zhu14-sh,wang2020intelligent,al2019online,rahman2013game,learning2020gamesec};
    ML-Based~\cite{nanda2016NFVSDN,badri2016uncovering,nisrine2016security,krueger2012ACMSAI,lee2010uncovering,lee2011seven,hofer2019model,ben2012combining,whitham2016minimising,stringhini2010detecting,whitham2017automating,abay2019using}
    
    & Honey files, honeypots, honeywebs, honeynets, honey patch, honey profiles, honey surface, honeybots, HMAC, obfuscation, deceptive signal, Fake Identities, deceptive network flow, social honeypots, bait-based deception, fake services & Zero-day attacks, APTs, masquerade attacks, reverse engineering for security patch, DoS attack, optimal inference attacks, reconnaissance attack, spam, social spam, malicious profiles & Signaling games, Stackelberg game, Metagames/Expected Utility, Bayesian Game using Bayesian equilibrium, subgame perfect Nash Equilibrium (SPNE), NLP techniques, hypergame, Stackelberg game, one-sided POSG, BayesNet, Decision Table, Na\"{i}ve-Bayes, Reinforcement Learning, SVM & Most game-theoretic DD approaches propose a general deception game framework without specifying a certain domain environment, which can have high applicability regardless of domain environments. & Due to the nature of a general approach, there is high overhead to conver the general approach to a domain-specific approach. \\
    \hline
    \new{General} Cyber-Physical Systems:
    GT-Based~\cite{kiekintveld2015game,sayin2019deception}; ML-based~\cite{hofer2019model}
    & Crafted bait information, honeypot & Reconnaissance attacks, node compromise, APT & Signaling game theory, RNN & Due to a wide range of CPS applications, developing DD techniques will have high values and applicabilities in diverse CPS environments. & DD techniques for the CPS do not really reflect unique challenges of the CPS environments. Physical honeypot deployment and maintenance often come with higher deployment and management cost. \\
    \hline
    Cloud web-based environments: GT-Based~\cite{pawlick2015flip,el2018new} & Honeywebs (using honeytokens, honey files, decoy resources, honeypots), deceptive signals, and obfuscation & General web attacks, APT attacks, privacy attacks & Signaling game theory, a suite of ML classifiers & Multiple DD frameworks are provided with high applicability to deal with a wide range of web attacks. & Some detailed designs should be considered to deal with unique challenges of cloud environments. \\
    \hline
    Internet-of-Things:
    GT-Based~\cite{la2016deceptive,xi2020hypergame,anwar2020gamesec,anwar2019game,anwar2020Honeypotbook,nan2019behavioral,anwar2020honeypot} & Honeypot & Reconnaissance and probing attacks, DoS attacks, packet dropping attacks, APTs & Bayesian games, meta games, signaling game with perfect Bayesian equilibrium & Since honeypots are fake nodes mimicking the behavior of a regular node, adding a honeypot \new{does not} change the hierarchy of the IoT network or the interface of IoT gateways. & A game-theoretic honeypot technique is the only technique that applied to this domain. However, the IoT naturally can generate a large amount of data for ML-based DD techniques which can create decoys that looks like real or enhance detect inside and outside attackers. \\
    \hline
    Software-defined networks:
    GT-Based~\cite{mao2019game,chiang2018defensive,al2019attacker,al2020dynamic}
    & Honeypot & Reconnaissance attacks & Bayesian game & The key advantage of using SDN-based defensive deception approaches is their easy deployability. For example, the existence of an SDN controller enables easily camouflaging a network topology or hiding vital nodes through flow traffic control. & Most existing approaches use a single SDN controller, which expose a single point of failure. \\ 
    \hline
    Wireless networks:
    GT-Based~\cite{clark2012deceptive, dimitriadis2007improving, nan2020mitigation} & Deceptive network flow, honeynet & Jamming attacks, 3G core network attacks  & A non-cooperative non-zero-sum static game & Specific design features to deal with key concerns of wireless networks are provided. & If a honeynet architecture is heavily dependent upon the accuracy of deployed honeypots, the formulated game may not work under different deployment settings.\\ 
    \hline
    \new{Online social networks: ML-Based~\cite{badri2016uncovering,lee2010uncovering,lee2011seven,stringhini2010detecting}} & \new{Social Honeypot, honey profiles} & \new{Fake liker (crowdturfing), social spammers, spamming} & \new{Supervised ML, ML toolkit, SVM} & \new{Learning the characteristics of attack behavior and attract spammers in OSN} & \new{Detectors are applied on static datasets to evaluate the detection of spammer and honey profiles are not generic and depend on the  platform type.} \\
    \hline
    \end{tabular}
\end{table*}

\subsection{Software-Defined Networks (SDNs)} \label{subsec:sdn}

The SDN paradigm separates data plane processing (e.g., packet forwarding) from control-plane processing (e.g., routing decisions)~\cite{MacFarland15}. The OpenFlow protocol~\cite{benton2013openflow} acts as an API between network switches and a logically centralized decision maker, called the {\em OpenFlow controller}. In this protocol, network switches cache data-plane flow rules. When a switch receives a packet and does not know how to forward it according to its cached rules, the switch sends an ``elevation'' request containing the original packet and a request for the guidance to the controller. The controller examines the packet and sends a set of rules that the switch should add to the data plane cache for forwarding packets~\cite{dixit2013towards}. deception techniques proposed in \cite{al2019attacker,al2020dynamic} were designed to secure SDN. 

\subsubsection{DD Techniques} 
Honeypots are popularly used as a defense strategy in game-theoretic defensive deception framework~\cite{mao2019game,chiang2018defensive} for taking dynamic, adaptive defense strategies. 

\subsubsection{Main Attacks} Common attack behaviors considered include scanning or reconnaissance attacks to obtain information and intelligence towards a target system by scanning network addresses (e.g., IP or port numbers) aiming to obtain defense information, network mapping, and inside information~\cite{chiang2018defensive} 

\subsubsection{Key GT and ML Methods}
Bayesian game theory has been popularly adopted under various conditions~\cite{mao2019game}, such as imperfect information, incomplete information, information sets, and perfect Bayesian equilibrium.

\subsubsection{Pros and Cons} The key advantage of using SDN-based defensive deception approaches is their easy deployability. For example, the existence of an SDN controller enables easily camouflaging a network topology or hiding vital nodes through flow traffic control. Most existing approaches use a single SDN controller, which exposes a single point of failure. 

\subsection{Wireless Networks}
\label{subsec:wireless-networks}

Wireless communications are everywhere these days and more common than wired communications due to their easy deployment and efficiency. However, when network resources are scarce, bandwidth constraints or unreliable wireless medium become main issues to be resolved. Game-theoretic or ML-based DD techniques are also proposed to defend against various types of attacks exploiting vulnerabilities of wireless network environments. 

\subsubsection{DD Techniques}
A deceptive network flow is proposed to generate the flow of random dummy packets in multihop wireless networks~\cite{clark2012deceptive}. A honeynet architecture address mobile network security (3G networks)~\cite{dimitriadis2007improving} was developed to enhance the security of the core network of mobile telecommunication systems. In \cite{dimitriadis2007improving}, a gateway was designed to control and capture network packets as well as investigate and protect other information systems from attacks launched from potentially compromised systems inside the honeynet. Defensive deception is used to transmit fake information over fake channels to mitigate jamming attacks in wireless networks~\cite{nan2020mitigation}.

\subsubsection{Main Attacks}
Game theoretic or ML-based DD techniques have defended against jamming attacks~\cite{clark2012deceptive} in multihop wireless networks and 3G core network attacks which compromise servers or gateways that support nodes providing general packet radio services in the mobile networks~\cite{dimitriadis2007improving}.

\subsubsection{Key GT and ML Methods}
A non-cooperative, non-zero-sum static game is used to model the interactions between an attacker and a defender~\cite{dimitriadis2007improving}. Deceptive routing paths are also designed based on a two-stage game using Stackelberg game theory~\cite{clark2012deceptive}. Deceptive power transmission allocation based on game theory is applied to defender wireless networks against jamming attacks \cite{nan2020mitigation}.

\subsubsection{Pros and Cons} Specific design features to deal with key concerns of wireless networks, such as multihop communications or mobile wireless security, are  helpful to implement real systems based on given game-theoretic DD technologies. However, like in~\cite{dimitriadis2007improving}, if a honeynet architecture heavily depends upon the accuracy of deployed honeypots, the formulated game framework may not guarantee the same level of effectiveness under different deployment settings.

\subsection{\new{Online Social Networks}} 
\new{
Due to the large scales of OSNs and their significant influence in social, economical, and political aspects, AI and ML communities have recognized high challenges in developing DD techniques in the OSN platforms.  Hence, there have been significant efforts made to secure OSNs from malicious users (e.g., fake users and spammers) by developing various social defensive deception techniques~\cite{badri2016uncovering,lee2010uncovering,lee2011seven, stringhini2010detecting}.
\subsubsection{DD Techniques} The authors in \cite{badri2016uncovering,lee2010uncovering,lee2011seven} specifically used social honeypots primarily to attract and identify attackers. Social honeypots are capable of learning the behavioural models of malicious users and collect a spammer's preferences. \citet{stringhini2010detecting} analyzed 900 honey profiles for detecting spammers across MySpace, Facebook, and Twitter for classification and identification purposes. 
\subsubsection{Main Attacks} OSNs are mainly targeted by fake likers and spammers that would like to exploit the existing dynamics of OSNs to achieve a specific goals~\cite{badri2016uncovering,lee2010uncovering,lee2011seven, stringhini2010detecting}. 
\subsubsection{Key GT and ML Methods}
Traditional ML tool kits such as SVM
have been used to analyze the data collected by social honeypots and honey profiles.
\subsubsection{Pros and Cons} Social honeypots allows learning the characteristics of attack behavior and honey files can attract spammers to protect real files from them in OSN. However, detectors are applied on static datasets to evaluate the detection of spammers. In addition, honey profiles are not generic and should be customized to  a specific platform type.
}

For the convenience of easy look-up for game-theoretic or ML-based DD techniques based on different application domains, we summarized our discussions of this section in Tables~\ref{tab:gt-summary} and~\ref{tab:ml-summary}. 

\section{Evaluation of Defensive Deception Techniques: Metrics and Testbeds} \label{sec:evaluation}

In this section, we survey what types of metrics are used to measure the effectiveness and efficiency of defensive deception techniques. In addition, we address what types of testbeds are employed to evaluate the effectiveness and efficiency of defensive deception (DD) techniques.

\subsection{Metrics} \label{subsec:metrics}
In the literature, the following metrics are used to measure the {\bf effectiveness} of existing defensive deception techniques:
\begin{itemize}[leftmargin=*]
\item {\em Detection accuracy}~\cite{pawlick2019game,Xu12-obfuscation,pawlick2018modeling,basak2019identifying, chiang2018defensive,badri2016uncovering,nisrine2016security,lee2010uncovering,lee2011seven, ben2012combining,abay2019using,yang2014taste}: This is often measured by the AUC (Area Under the Curve) of ROC (Receiver Operating Characteristic). AUC is used to measure the accuracy of detection by showing TPR (True Positive Rate) as FPR (False Positive Rate) increases. \citet{abay2019using} used a classifier accuracy as a metric to evaluate the effectiveness of the developed Honey data to deceive the attacker. \citet{badri2016uncovering} and \citet{chiang2018defensive} used an algorithm to discover fake Liker in social networks. The authors in \cite{Xu12-obfuscation,ben2012combining} evaluated a masquerade attack detector based on AUC. ROC is also used to measure detection accuracy of social honeypot-based approach against social spammers~\cite{lee2010uncovering,lee2011seven} and malicious account~\cite{nisrine2016security,yang2014taste} in online social networks. \citet{pawlick2019game,pawlick2018modeling} evaluated their deception mechanism based on different metrics, including the detector accuracy developed by the adversary to discover deception.

\item {\em Mean time to detect attacks}~\cite{basak2019identifying}: The effectiveness of a detection mechanism is also measured by how early their deception technique can capture the attacker.

\item {\em Utility}~\cite{pawlick2019game,kiekintveld2015game,clark2012deceptive,zhu2012deceptive,mohammadi2016Springer,Zhang19-subgame,pawlick2018modeling,pibil2012game,xi2020hypergame,Stephanie2020Harnessing,Olivier2020Partially,anwar2020gamesec,anwar2019game,anwar2020Honeypotbook,nan2019behavioral,nan2020development, garg2007deception,horak2017manipulating,huang2019dynamic,casey2016compliance,thakoor2019cyber,nan2020mitigation,wang2020intelligent,anwar2020honeypot}: In game-theoretic DD techniques, a player (either an attacker or a defender)'s utility (or payoff) is considered as one of key metrics to evaluate a deception game between the attacker and defender. Commonly, the utility of taking a defensive deception strategy is formulated based on the deployment cost and the confusion increased to an attacker. Some signaling games model the interactions between an attacker, a benign client and a defender. As a result, some of these studies considered the impact to normal as another cost of deception technology~\cite{wang2020intelligent,rahman2013game}. A defender's expected utility is a common metric to be maximized as the system's objective, as shown in game-theoretic or ML-based DD techniques in \cite{pawlick2019game,kiekintveld2015game,zhu2012deceptive,Zhang19-subgame,pawlick2018modeling,pibil2012game, xi2020hypergame,Stephanie2020Harnessing, Olivier2020Partially,nan2020development,garg2007deception, horak2017manipulating,huang2019dynamic,thakoor2019cyber,al2019attacker,al2020dynamic}. 
\item {\em Probabilities of an attacker taking certain actions}~\cite{wagener2009self,nan2019behavioral,bilinski2019you,rahman2013game,cranford2020adaptive, cranford2019towards}: An attacker's actions (i.e., choices) in proceeding attacks are also considered as a metric to measure the effectiveness of defensive deception (e.g., honeypots). For example, in~\cite{wagener2009self}, the following probabilities of an attacker's action are considered: a probability of an attacker retrying a failure command, a probability of an attacker choosing an alternative strategy, and a probability of an attacker leaving a game. The probability of an attacker to successfully control the network and overcome the deception is used to evaluate the effectiveness of cyberdeception in~\cite{bilinski2019you, rahman2013game} as well as the probability of launching an attack as in~\cite{cranford2020adaptive, cranford2019towards}. 
\end{itemize}

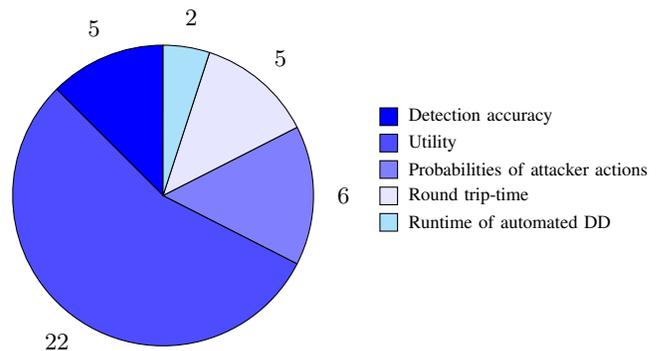
\begin{figure}[htb]
\vspace{-3mm}
\centering
\begin{tikzpicture}
[pie chart,
slice type={Detection accuracy}{blue},
slice type={Utility}{blue!70},
slice type={Probabilities of attacker actions}{blue!50},
slice type={Round trip-time}{blue!10},
slice type={Runtime of automated DD}{cyan!30},
pie values/.style={font={\small}},
scale=2]
\pye[xshift=0.5cm, values of coltello/.style={pos=1}, radius =0.5]{1994}
{
5/Detection accuracy,
22/Utility,
6/Probabilities of attacker actions,
5/Round trip-time,
2/Runtime of automated DD}
\legend[shift={(2cm, 0.7cm)}]{{Detection accuracy}/Detection accuracy, {Utility}/Utility, {Probabilities of attacker actions}/Probabilities of attacker actions, {Round trip-time}/Round trip-time, {Runtime of automated DD}/Runtime of automated DD}
\end{tikzpicture}
\caption{Types and frequency of metrics measuring performance and security of game-theoretic defensive deception techniques.}
\label{fig:metrics-GT}
\end{figure}

\begin{figure}[htb]
\vspace{-3mm}
\centering
\begin{tikzpicture}
[pie chart,
slice type={Detection accuracy}{blue},
slice type={Utility}{blue!70},
slice type={Round trip-time}{blue!10},
slice type={Runtime time}{cyan!30},
pie values/.style={font={\small}},
scale=2]
\pye[xshift=0.5cm, values of coltello/.style={pos=1}, radius =3]{1994}
{
8/Detection accuracy,
3/Utility,
2/Round trip-time,
1/Runtime time}
\legend[shift={(2cm, 0.7cm)}]{{Detection accuracy}/Detection accuracy, {Utility}/Utility, {Round trip-time}/Round trip-time,{Runtime time}/Runtime time}
\end{tikzpicture}
\caption{Types and frequency of metrics measuring performance and security of ML-based defensive deception techniques.}
\label{fig:metrics-ML}
\vspace{-3mm}
\end{figure}
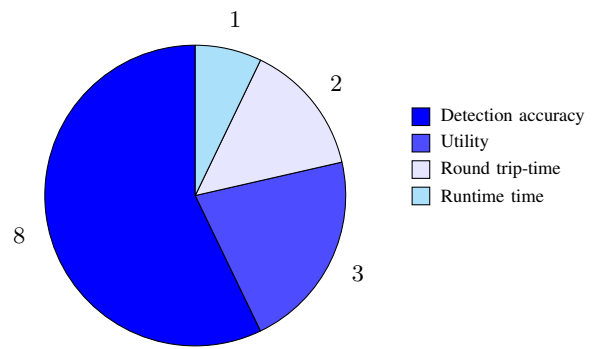

In addition, we also found the following metrics to capture the {\bf efficiency} of defensive deception (DD) techniques discussed in our survey paper:
\begin{itemize}[leftmargin=*]
\item {\em Round trip-time}~\cite{Borello08, Chan04-jss,  basak2019identifying,Araujo14,aggarwal2019hackit,  cai2009attacker}:  The key role of DD is to delay the adversary processes~\cite{Araujo14}. Increasing the time required to crack an obfuscated code is the evaluation metric adopted in~\cite{Chan04-jss,cai2009attacker}, as well as the complexity of the deobfuscation process~\cite{Borello08}. The HackIt deception tool in~\cite{aggarwal2019hackit} successfully increased the time taken by hackers to exploit a system. An attacker's deception detection time is of great significance to evaluate the effectiveness of cyber deception as in~\cite{basak2019identifying}. 

\item{\em Runtime}~\cite{Avery17, avery2018CS,learning2020gamesec}: 
This evaluates the cost of automated deception and obfuscation techniques. \citet{Avery17} and \citet{avery2018CS} evaluated the performance of the developed fake patches deception technique based on this runtime metric. Similarly, the learning time is the metric used to evaluate the learning the attacker’s behavior model from attack data introduced in \cite{learning2020gamesec}.
\end{itemize}
In Figs.~\ref{fig:metrics-GT} and~\ref{fig:metrics-ML}, we summarized the types and frequency of metrics measuring performance and security of the DD techniques surveyed in this work. As shown in Fig.~\ref{fig:metrics-GT}, a player's utility is the most common metric among all metrics used for the surveyed game-theoretic DD techniques. 
On the other hand, in Fig.~\ref{fig:metrics-ML}, ML-based DD approaches heavily relied on detection accuracy using such metrics as AUC, TPR, and FPR metrics. 

\subsection{Evaluation Testbeds}
In this section, we classify evaluation testbeds of the existing game theoretic or ML-based DD techniques surveyed in this work in terms of the four classes: probability-based, simulation-based, emulation-based, and real testbed-based. We discuss each of the evaluation testbeds and discuss the general trends observed from the survey.
 
\subsubsection{Probability Model-Based Evaluation~\cite{Cho19-hgt,cho18-jdms}}

Stochastic Petri Nets (SPN) probability models have been used to evaluate the performance of an integrated defense system that combines multiple deceptive defense techniques~\cite{cho18-jdms}. In addition, the SPN is used to measure the performance of the deceptive defender and the attacker in a hypergame model~\cite{Cho19-hgt}.

\subsubsection{Simulation Model-Based Evaluation~\cite{pawlick2019game,Borello08, Xu12-obfuscation,kiekintveld2015game,clark2012deceptive,mohammadi2016Springer,House10,Zhang19-subgame,pawlick2018modeling,  pibil2012game,la2016deceptive, basak2019identifying,xi2020hypergame,Kulkarni20, Stephanie2020Harnessing,Olivier2020Partially,anwar2020gamesec, anwar2019game,anwar2020Honeypotbook,nan2019behavioral,nan2020development, garg2007deception,yin2013optimal, horak2017manipulating,bilinski2019you,chiang2018defensive, huang2019dynamic,thakoor2019cyber,Avery17, cadar2008klee,al2019attacker, al2020dynamic,nan2020mitigation, anwar2020honeypot,al2019online,rahman2013game,learning2020gamesec,al2019towards,badri2016uncovering, ben2012combining, abay2019using,chiang2016acyds,chadha2016cybervan}}  In order to simulate benign network traffic, \citet{rahman2013game} generated network flow by the Internet Traffic Archive~\cite{Danzig08}, using {\em Nmap} to test their deception technology. ~\cite{chiang2018defensive,chiang2016acyds} to simulate their deception system and combine it with their proposed work. \citet{thakoor2019cyber} used the CyberVan~\cite{chadha2016cybervan,pham2020quantitative} as a network simulation testbed to simulate their game-theoretic cyberdeception techniques. \citet{al2019online} and \citet{bilinski2019you} proposed a deceptive system which was simulated using a Markov decision processes. Several studies also simulated ML-based deceptive algorithms or classifiers to examine its accuracy~\cite{ al2019attacker,al2020dynamic,al2019online,learning2020gamesec,al2019towards,badri2016uncovering,ben2012combining,abay2019using}. A masquerade attack is simulated to evaluate the developed detection techniques based on deception~\cite{Xu12-obfuscation,ben2012combining}. Various types of game-theoretic based deception framework have been evaluated based on simulation models~\cite{kiekintveld2015game,clark2012deceptive,mohammadi2016Springer,House10,Zhang19-subgame,pawlick2018modeling,pibil2012game,la2016deceptive,basak2019identifying,xi2020hypergame,Kulkarni20,Stephanie2020Harnessing,Olivier2020Partially,anwar2020gamesec,anwar2019game,anwar2020Honeypotbook,nan2019behavioral,nan2020development,garg2007deception,yin2013optimal,horak2017manipulating,bilinski2019you,chiang2018defensive,huang2019dynamic,thakoor2019cyber,Avery17,cadar2008klee,anwar2020honeypot,rahman2013game}. Simulation-based models showing a proof of concept have been developed to demonstrate the runtime of the proposed automated deception algorithm~\cite{Avery17, cadar2008klee}. 

\subsubsection{Emulation Testbed-Based Evaluation~\cite{Chan04-jss, wang2020intelligent, dixit2013towards}}

Emulation network environments have been proposed to dynamically place deception resources~\cite{wang2020intelligent}, deploy code obfuscation techniques on real-like decompilers~\cite{Chan04-jss}, and to deploy DD techniques in emulated SDN-based data center network testbed. 

\subsubsection{Real Testbed-Based Evaluation~\cite{Ardagna07-location-obfuscation,aggarwal2016cyber, aggarwal2017modeling, Araujo14, aggarwal2019hackit,  nisrine2016security, krueger2012ACMSAI, lee2010uncovering, lee2011seven, yang2014taste}} 
An application for data collection has been developed~\cite{aggarwal2016cyber, aggarwal2017modeling, aggarwal2019hackit}. Experiments with human-in-the-loop with human participants (e.g. an attacker or a defender) have been conducted~\cite{aggarwal2016cyber, aggarwal2017modeling, aggarwal2019hackit}, such as HackIt to evaluate the effectiveness of deception on participating hackers under different deception scenarios. 
Socialbots on Twitter real networks are significantly leveraged to detect and study content polluters and understand how spammers choose their spamming targets to ultimately create a malicious profile classifier~\cite{nisrine2016security,lee2010uncovering,lee2011seven, yang2014taste}. In addition, various DD techniques are validated under real settings, such as honey-patches-based deception system~\cite{Araujo14}, location obfuscation for preserving users' privacy~\cite{Ardagna07-location-obfuscation}, and PRISMA (PRotocol Inspection and State Machine Analysis) tool capable of learning and generating communication sessions for deception~\cite{krueger2012ACMSAI}. 

\begin{figure}
    \centering
\begin{tikzpicture}
[pie chart,
slice type={Simulation model-based}{blue!50},
slice type={Emulation model-based}{blue!10},
slice type={Real testbed-based}{cyan!30},
pie values/.style={font={\small}},
scale=2]
\pye[xshift=0.5cm, values of coltello/.style={pos=1}, radius =0.3]{1994}
{
8/Simulation model-based,
1/Emulation model-based,
5/Real testbed-based}
\legend[shift={(2cm, 0.5cm)}]{ {Simulation model-based}/Simulation model-based, {Emulation model-based}/Emulation model-based, {Real testbed-based}/Real testbed-based}
\end{tikzpicture}
\caption{Evaluation testbeds for ML-based defensive deception techniques.}
\label{fig:testbeds-ml}
\end{figure}
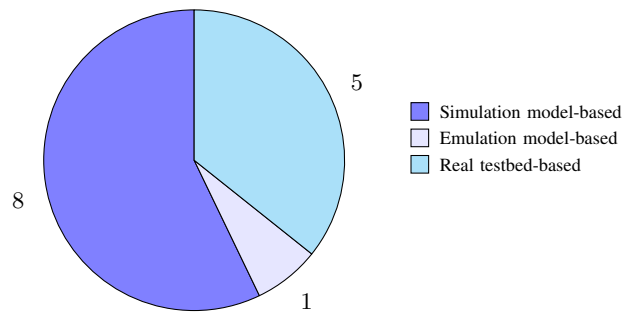

In Figs.~\ref{fig:testbeds-GT} and~\ref{fig:testbeds-ml}, we summarized the frequency of particular testbeds used in developing game-theoretic and ML-based defensive deception techniques surveyed in this paper, respectively.  The overall trends observed were that simulation-based evaluation was dominant in both GT and ML-based DD approaches. Although ML-based DD techniques are less explored than GT-based DD techniques, it is noticeable that more than 40 percent of ML-based DD techniques surveyed in this paper used the real testbed-based evaluation. This is natural that ML-based approaches basically need actual datasets to analyze. Another noticeable finding is that there has been a lack of probability model-based approaches. This would be because it is highly challenging to model a complex system environment characterized by many design parameters based on mathematical models.

\begin{figure}
    \centering
\begin{tikzpicture}
[pie chart,
slice type={Probability model-based}{blue!70},
slice type={Simulation model-based}{blue!50},
slice type={Real testbed-based}{cyan!30},
pie values/.style={font={\small}},
scale=2]
\pye[xshift=0.5cm, values of coltello/.style={pos=1}, radius =0.3]{1994}
{
1/Probability model-based,
31/Simulation model-based,
4/Real testbed-based}
\legend[shift={(2cm, 0.7cm)}]{{Probability model-based}/Probability model-based, {Simulation model-based}/Simulation model-based,  {Real testbed-based}/Real testbed-based}
\end{tikzpicture}
\caption{Evaluation testbeds for game-theoretic defensive deception techniques.}
\label{fig:testbeds-GT}
\end{figure}
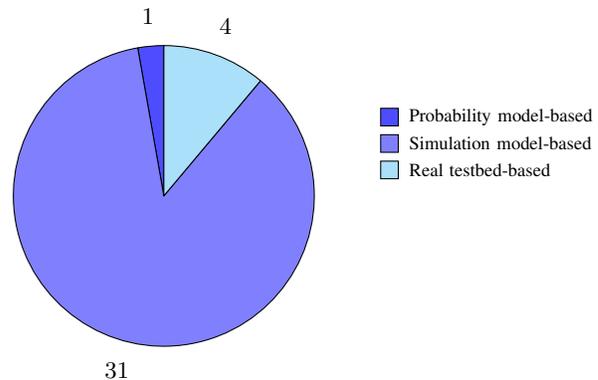

\section{Conclusions and Future Work}
\label{sec:conclusions}
In this section, we discussed insights and lessons learned and limitations from the extensive survey conducted in this work.  In addition, based on the conducted survey, we provided a set of promising future research directions.

\subsection{Insights and Lessons Learned}
\label{subsec:insights-lessons}

We now revisit and answer the research questions raised in Section~\ref{subsec:research-questions}.

\begin{description}[leftmargin=3mm]
\item [RQ Characteristics:] {\em What key characteristics of defensive deception distinguish it from other defensive techniques?}

\vspace{1mm}
{\bf Answer}:  Unlike traditional defense mechanisms, deception requires a certain level of risk as it requires some interactions with attackers with the aim of confusing or misleading them.  Particularly, if the goal of defense requires long term deception, it is inevitable to face risk. In this case, defensive deception can be used with other legacy defense mechanisms such as intrusion prevention or detection mechanisms to avoid too high risk.  Moving target defense (MTD) or obfuscation techniques share a similar defense goal with defensive deception, such as increasing confusion or uncertainty for attackers. However, unlike MTD or obfuscation which changes system configuration or information based on the existing resources of a system, defensive deception can create false objects or information with the aim of misleading an attacker's cognitive perception or forming a misbelief for the attacker to choose a sub-optimal or poor attack strategy. 

\vspace{1mm}
\item [RQ Metrics:] {\em What metrics are more or less used to measure the effectiveness and efficiency of the existing game-theoretic or ML-based defensive deception techniques?}

\vspace{1mm}
{\bf Answer}: As surveyed in Section~\ref{sec:evaluation} along with Figs.~\ref{fig:metrics-GT} and~\ref{fig:metrics-ML}, the effectiveness of GT-based deception is mainly measured based on utility.  On the other hand, ML-based deception is mainly evaluated according to the accuracy of classifiers or intrusion detectors. \new{The existing metrics observed in the surveyed literature for game-theoretic or ML-based DD approaches do not capture the direct impact of defensive deception. That is, they do not quantify the uncertainty or confusion induced by the different deceptive techniques. Moreover, new metrics should be developed to measure the effectiveness of deception in misleading the attackers, such as lightweight deception with less or minimum defense cost in deploying or maintaining a given deception technique.}

\vspace{1mm}
\new{According to Section~\ref{sec:evaluation} along with Figs.~\ref{fig:metrics-GT} and~\ref{fig:metrics-ML}, GT-based deception techniques have been mainly evaluated in terms of the defender's utility. However, ML-based deception techniques focused on improving the accuracy of classifiers or intrusion detectors. However, the existing evaluation of GT-based defensive deception techniques rarely consider utility analyses based on the losses and gains in terms of the timing of using different types of deceptions. In addition, although it is important to use metrics capturing deception cost introduced by its deployment and maintenance and performance degradation caused by running deception techniques, there has been much less effort to use those metrics.  Furthermore, measuring the extent of attackers deviating from the optimal course of actions can be another meaningful metric to evaluate the effectiveness of deception in attack-defense games.}
\vspace{1mm}

\item [RQ Principles:] {\em What key design principles help maximize the effectiveness and efficiency of defensive deception techniques?}

\vspace{1mm}
{\bf Answer}: It is critical to make three key design decisions, which are what-attacker-to-deceive, when-to-deceive, and how-to-deceive. These three design decisions should be determined based on critical tradeoffs between effectiveness and efficiency of a developed DD technique. The what-attacker-to-deceive question should be answered based on what attacks are targeted by a defender and what attacks should be commonly handled in a given application domain because each application domain generates distinctive challenges. \new{When-to-deceive is related to the fact that depending on when a given defensive deception technique is used to deal with attackers in a certain stage of the cyber kill chain, the extent of deployability, effectiveness, efficiency, and cost of using a given DD technique can be affected.  For example, when-to-deceive can be decided according to how complex the deployment of such DD technique or how effective the DD technique is in terms of the timing. In addition, it can consider how costly deception is in terms of the availability of the resources.  In some settings, deception is needed only temporarily to delay an attack until a defense system needs to take enough time for its reconfiguration or deploying/performing a certain defense operation (e.g., deceive an attacker until the MTD is executed to fully reconfigure an OS). Each type of a DD technique has its own unique characteristics and corresponding expected outcomes. Given what attack to deal with by determining what-to-deceive and when to deceive the attacker (i.e., in the attack stage), the how-to-deceive question is related to answering (1) what DD techniques are better than other DD techniques under given resources and (2) what quality of deception is most appropriate to deal with the given attacks. For example, if a given defense system can afford to deploy high-quality honeypots to deal with highly intelligent attackers, high-interaction honeypots would be a good choice. However, if the defense system has a limited budget or resources available and wants to deploy a lightweight DD technique temporarily (e.g., before a new setting of MTD is fully reconfigured), a honey file or honey token can be used with low cost but they can be detected by intelligent attackers soon.  Therefore, depending on the defense cost budget, the system's resource availability, or the expected outcome for a DD technique to be used in a given time, a different type of DD techniques with a different level of deception quality can be adopted to meet the expected effectiveness and efficiency.} 
\vspace{1mm}

\item [RQ GT:] {\em What are the key design features when a defensive deception technique is devised using game theory?}

{\bf Answer}: \new{Game-theoretic DD techniques mainly adopt an attack-defense framework where the attacker and defender interact with a conflict of interest scenario, which often use a form of zero-sum games. However, as multiple players in each party (i.e., multiple attackers for collusive attacks or multiple defenders for cooperative defense) can involve in a given attack-defense game, general-sum games are often considered as well.  To be more specific, we also summarized a game type used in each paper that proposes a game theoretic defensive deception in Table~\ref{tab:gt-summary}.} Game-theoretic DD focuses on identifying optimal defense strategies to confuse or mislead an attacker by increasing uncertainty. Three design features are: (i) each player needs to have a clear goal and a corresponding clear utility function that reflects the player's intent, tactics, and resources; (ii) modeling and simulating how to increase confusion or uncertainty via different types of DD techniques should be elaborated; and (iii) metrics to measure effectiveness and efficiency should be articulated.

\vspace{1mm}

\item [RQ ML:] {\em What are the key design features when a defensive deception technique is developed using ML?}

{\bf Answer}: ML-based DD techniques learn and detect attack behaviors. However, ML can improve DD by considering the following: (i) We need to consider what types of datasets use to develop DD techniques. To develop believable fake objects presupposes good datasets for mimicking real objects and evaluating fake objects. Along these lines, targeted attack datasets should be used to model attackers and develop honey resources; (ii) ML-based DD approaches should adopt relevant metrics to capture their effectiveness and efficiency. So far, classification accuracy is the only metric to capture ML-based honeypots. Additional metrics should be developed to capture the quality of ML-based DD techniques; and \new{(iii) A fully automated deception is highly desirable for ML-based DD techniques, such as generating deceptive traffics and network topologies.}

\vspace{1mm}

\item [RQ Applications:] {\em How should different defensive deception techniques be applied in different application domains?}

As the enterprise network is a complex domain, multiple honey-X techniques, possibly in combination, are applicable. For simpler domains, an easy-to-deploy honeypot is effective and widely used. \new{However, developing high quality defensive deception techniques often incurs more cost or higher complexity.} For example, deploying more honeypots may increase the chance to mislead an attacker. However, too many honeypots also increase unnecessary costs and confuse the administrator or legitimate users.  Depending on applications, DD techniques should be applied in an automated manner. For example, online social networks require deception for detecting attacks or collecting attack intelligence since social spammers do not aim to gain control but to influence other users in the network. 

\end{description}

\new{As you can observe in our extensive survey, many more game theoretic defensive deception studies have been conducted while ML-based DD studies are limited to developing honey-X-based approaches aiming to detect more attacks. Via our extensive survey, we could clearly observe the main benefit of using game theoretic DD techniques is their capability to allow players to make strategic decisions and take various strategies upon observed dynamics of a given game over time. In addition, since game theory provides a mathematical framework to model the interactions between attackers and defenders, it can be more extensively applied in developing various types of DD techniques as a general framework. However, as we also observed in existing work, although game theoretic DD has powerful capability with solid mathematical proofs in developing DD techniques, their validation in security and performance has been heavily limited to the theoretical or simulation-based verification. However, ML-based DD approaches have been applied to mainly honey-X applications, mostly developing honeypots.  Unlike other DD techniques, the main goal of honeypots are two-fold: protecting system assets by creating fake objects, which are honeypots and detecting attackers by luring them to the honeypots. Although the original purpose of DD more focuses on protecting the system assets, which is more like passive defense, the effectiveness of honeypots has been mainly measured its role in detecting attacks in terms of the detection accuracy. That is, ML-based honeypots have been developed by creating a fake object which looks like real with the aim of attracting more attackers. Compared to GT-based DD approaches, ML-based DD approaches are less dynamic because ML techniques are mainly used to improve the quality of deception in its development time and as a classifier to detect attacks.  Although reinforcement learning as one of promising ML techniques as been used to develop a DD technique along with game theory~\cite{wang2020intelligent,al2019online}, hybrid approaches incorporating ML into game theory to develop DD techniques are rarely taken. We discuss examples of how these can be incorporated into a hybrid DD technique in the future research directions discussed in Section~\ref{subsec:future-research}.}

\subsection{Limitations of Current Defensive Deception}
\label{subsec:limitations}
From our extensive survey on the state-of-the-art defensive deception (DD) techniques using game theory and ML, we found the following limitations of the existing approaches:
\begin{itemize}
\item Although many DD techniques have used game-theoretic approaches, they have mainly focused on theoretical validation based on Nash equilibrium and the identification of optimal solutions, in which attacks are not often detailed enough but simply considered attacks compromising other nodes. Often times attacks are complicated and performed with more than one episode or multistage over time, such as APT attacks. Therefore, although we can benefit from modeling simple attack processes, such as active reconnaissance or insider attack, it is still challenging to deploy the game-theoretic DD in real systems and model attacks based on a complex cyber kill chain.

\item We observed that an enterprise networking is the only domain that is well developed. Research effort in other domains has been limited and mainly focused on reconnaissance attacks. IoT and SDN environments are capable of generating large amounts of traffic flow data, which can be used to train ML models to identify attackers. However, current DD approaches for those domains lack using ML.

\item Due to high complexity of action spaces, most game-theoretic DD has dealt with a limited set of actions for an attacker and defender, which cannot accurately model an attack or defensive process when the attacker can perform multiple attacks following the multiple stages of attack processes, such as cyber kill chain in APT attacks.

\item When multiple attacks arrive in a system, considering the interactions only between one attacker and one defender is not sufficient to capture the practical challenge. How to interpret a response towards a certain action (e.g., who is responding to what action when there are multiple attackers) is not clear although it is critical to model the interactions between the attacker and defender.

\item Majority of game-theoretic defensive deception approaches are studied at an abstract level by using game-theoretical analysis without explicitly addressing design challenges derived from the network environment or platform the defensive deception technique is deployed. When some ideas are adopted and applied in a certain network environment, it is quite challenging how to deploy the given game-theoretic work into a given platform due to lack of details available for the deployment process.

\item The accuracy of ML-based deception techniques depends on data availability regarding the attacker identity, techniques, and targets. Practically, such data are not available for the defender, which significantly limits training ML-based classifiers or detectors. Moreover, such models tend to assume that the attacker is acting normally toward its target. However, if an attacker decided to deceive a defender to remain stealthy, the effectiveness of defensive deception techniques may not be clearly measured.
\item Compared to game-theoretic defensive deception techniques, ML-based defensive deception has been much less studied. The majority of using ML techniques in defensive deception is to detect attacks in honeypots. Only a few works have used ML to mimic real objects or information for developing honey objects or information.
\item \new{Although game-theoretic or ML-based defensive deception has been studied, there have been a few studies that combine both to develop hybrid defensive deception techniques, particularly incorporating reinforcement learning (RL) into players' utility functions and using RL to identify the players' optimal strategies.}
\item Evaluation metrics of defensive deception are limited. Most game-theoretic DD approaches mainly used metrics in game theory, such as utility or probability of taking attack strategies. Most ML-based DD approaches were mainly studied to achieve attack detection, which leads to using detection accuracy as a major metric. These metrics, utility and detection accuracy, do not fully capture the effectiveness and efficiency of DD approaches based on the core merit of using deception.
\item Evaluation testbeds are mostly based on simulation models. Although some ML-based DD approaches used real testbeds, they are mostly social honeypots deployed in social media platforms. In particular, game-theoretic DD approaches remain highly theoretical analysis.
\end{itemize}

\subsection{Future Research Directions}
\label{subsec:future-research}

We suggest the following research directions as promising future research:
\begin{itemize}
\item \new{{\bf Need more metrics to measure the quality of honey-X technologies}: The quality of honey-X has been captured mainly based on detection accuracy (see Figs.~\ref{fig:metrics-GT} and \ref{fig:metrics-ML}) even if the key role of honey-X is to protect assets as well as to detect attacks. Hence, there should be more diverse metrics to measure the roles of both protecting assets and detecting attacks. The example metrics can include an attacker's perceive uncertainty level, the amount of missing vulnerable assets by attackers, the amount of novel attack vectors collected via honeypots, or the amount of important system components attacked.  In addition, the efficiency of running honey-X technologies (i.e., defense cost) or performance degradation due to the deception deployment should be considered for a defender to choose an optimal strategy based on multiple criteria including those multiple metrics as objectives.}
 
\item \new{{\bf Consider more sophisticated, intelligent attackers}: The attacker behaviors in the literature mostly consider simple attacks which are not more intelligent than the defender behaviors, which are not realistic in practice. In particular, a highly intelligent attacker may be able to detect the deception deployed by the defender. Further, in most existing game theoretic approaches using Stackelberg games, the defender is a leader while the attacker is a follower where there is a first-mover advantage in such sequential games. This may not be true in real-world cyber settings. In addition, when APT attacks are considered, most existing approaches consider only reconnaissance attacks. In future work, we need to consider the APT attackers performing multistaged attacks following the cyber kill chain (CKC) stages (i.e., including reconnaissance, delivery, exploitation, command and control (C2), lateral movement, and data exfiltration).}
\item \new{{\bf Measure the quality of defensive deception}: The quality of a defensive deception technique should be determined based on how well it deceives attackers.  That is, the quality of defensive deception should be measured by the attacker's view and actions based on its belief towards the defender's moves. However, existing work mainly uses system metrics as a proxy to measure the quality of a defensive deception technique deployed by the defender. Some example metrics can be the degree of uncertainty perceived by attackers, the degree of the discrepancy between what the attackers perceived towards the defense system and the ground truth system states, or service availability without being disrupted by deployed defensive deception techniques.}

\item \new{{\bf Develop hybrid defensive deception techniques based on both game theory and ML}: Machine learning-based game theoretic approaches (e.g., game theory plus deep reinforcement learning) have been considered in developing other defense techniques~\cite{Liu20-iot-drl, Xu20-game-rl-msn-tifs}. However, only a few studies~\cite{wang2020intelligent,al2019online} used hybrid approaches of incorporating reinforcement learning (RL) with game theory where we treat RL as one of ML techniques. RL's reward functions can be used to formulate players' utility functions and let an RL agent identify an optimal strategy as other attack-defense games~\cite{Liu20-iot-drl, Xu20-game-rl-msn-tifs} leverage RL or deep RL (DRL). In addition, ML can be used for players to estimate their own beliefs to decide what strategy to take or opponents' beliefs to predict their moves.  In addition, when players use a game-theoretic approach to take an optimal strategy, a set of DD techniques can be developed using ML technologies to increase the quality of the deception by mimicking real objects or information.}

\item \new{{\bf Measure cost or service availability under the deployment of defensive deception techniques}: As discussed in Section~IX-B, most game-theoretic DD studies used utility and probabilities of taking strategies while ML-based DD approaches mainly used detection accuracy as the major metric.  However, the drawbacks introduced by deploying defensive deception techniques have not been sufficiently discussed. In addition, some strong assumptions, such as no impact introduced to the defender system, are made while deploying defensive deception techniques. By considering metrics to measure performance degradation introduced by a used defensive deception, we can measure a broader aspect of the quality of the defensive deception deployed in a system based on the classical tradeoff between security and performance.} 

\item \new{{\bf Build a realistic testbed to evaluate game theoretic defensive deception techniques}: Game-theoretic DD approaches have been dominantly studied based on the theoretical validation via mathematical proof or experimental analysis via simulation models. More realistic emulation or real testbeds should be reflected to validate game theoretic DD approaches considering highly intelligent attackers and highly uncertain network environments. To this end, there should be clear designs on how to deal with uncertainty in terms of how much knowledge is known for an attacker, what system information (i.e., a defender's information) can be known to the attacker, and what network environment information or dynamics are known to both the defender and the attacker. How much and how accurate information is available to players and how their beliefs are formulated under real settings are critical points to tackle because strong assumptions on common knowledge in game theory cannot hold in real environments. The first step to tackle this problem is to build a game where two players may have asynchronous information with imperfect or incomplete information, such as a hypergame~\cite{bakker2021metagames}.}

\item \new{{\bf Develop a domain-specific game theoretic framework that can provide defensive deception as a security service}: Based on the observation of Table~\ref{tab:gt-summary}, we find many game theoretic frameworks are developed as general frameworks without specifying an domain application. This would be helpful for general understanding about the core idea of the method itself. However, it misses how to design defensive deception techniques for a specific domain environment which may pose challenges in terms of resources available (e.g., bandwidth, computational power), environmental dynamics, or other characteristics of common adversarial attacks. Providing more specific applications and corresponding system designs for developing and deploying defensive deception techniques will lead to their improved applicability into the real environments.}

\item \new{{\bf Identify highly compatible configurations of using both defensive deception and legacy defense services}: It is not necessarily true to always use defensive deception techniques to handle all types of attacks. In addition, it is not always true that DD techniques should be always deployed with legacy defense services, such as intrusion prevention, detection, and response systems. This is because some combinations of deploying defensive deception techniques with legacy defense services, such as using honeypots with intrusion detection and response systems, may introduce overlapping effect which is not really efficient (e.g., both honeypots and IDS may do a same job).  We should develop a more systematic method to figure out how to synergistically leverage both defense services to provide cost-effective defense services.}

\end{itemize}

\section*{Acknowledgements}
This research was partly sponsored by the Army Research Laboratory and was accomplished under Cooperative Agreement Number W911NF-19-2-0150. In addition, this research is also partly supported by the Army Research Office under Grant Contract Numbers W91NF-20-2-0140 and W911NF-17-1-0370. The views and conclusions contained in this document are those of the authors and should not be interpreted as representing the official policies, either expressed or implied, of the Army Research Laboratory or the U.S. Government. The U.S. Government is authorized to reproduce and distribute reprints for Government purposes notwithstanding any copyright notation herein.

\bibliographystyle{IEEEtranN}

\bibliography{ref}

\end{document}